\documentclass[aps,twocolumn,prl, superscriptaddress,amsmath,amssymb,showpacs]{revtex4-1}
\usepackage{graphicx}
\usepackage{amsmath}
\usepackage{comment}
\usepackage{bm}
\usepackage{color}

\renewcommand{\thesection}{\Roman{section}}

\begin{document}

\title{Multiple-$q$ states of the $J_1$-$J_2$ classical honeycomb-lattice Heisenberg antiferromagnet under magnetic fields}

\author{Tokuro Shimokawa} 
\email{tokuro.shimokawa@oist.jp}
\affiliation{Okinawa Institute of Science and Technology Graduate University, Onna, Okinawa, 904-0495, Japan}
\author{Tsuyoshi Okubo}
\affiliation{Department of Physics, University of Tokyo, Tokyo 113-0033, Japan }
\author{Hikaru Kawamura}
\affiliation{Department of Earth and Space Science, Graduate School of Science, Osaka University, Toyonaka, Osaka 560-0043, Japan}

\date{\today}

\begin{abstract}
 Motivated by the recent theoretical study by Okubo $et \ al$ [Phys.~Rev.~Lett.~${\bf 108}$,~017206~(2012)] on the possible realization of the frustration-induced {\it symmetric\/} skyrmion-lattice state in the $J_1$-$J_2$ (or $J_1$-$J_3$) triangular-lattice Heisenberg model without the Dzyaloshinskii-Moriya interaction, we investigate the ordering of the classical $J_1$-$J_2$ honeycomb-lattice Heisenberg antiferromagnet under magnetic fields by means of a Monte Carlo simulation, a mean-field analysis and a low-temperature expansion. The model has been known to have an infinite ring-like degeneracy in the wavevector space in its ground state for $1/6<J_2/J_1<0.5$, in distinction with the triangular-lattice model. As reported by Okumura $et \ al$ [J.~Phys.~Soc.~Jpn.~${\bf 79}$,~114705~(2010)], such a ring-like degeneracy gives rise to exotic spin liquid states in zero field, {\it e.g\/}, the ``ring-liquid" state and the ``pancake-liquid'' state. In this paper, we study the in-field ordering properties of the model paying attention to the possible appearance of exotic multiple-$q$ states. Main focus is made on the  $J_2/J_1=0.3$ case, where we observe a rich variety of multiple-$q$ states including the single-$q$, double-$q$ and triple-$q$ states. While the skyrmion-lattice triple-$q$ state observed in the triangular-lattice model is not realized, we instead observe an exotic double-$q$ state consisting of meron/antimeron-like lattice textures.
\end{abstract}

\maketitle

\section{\label{sec:Introduction}I. Introduction}
Frustrated spin systems have attracted much interest in the field of magnetism.  One of such research interest might be that novel types of ordering are often generated by the effects of quantum or thermal fluctuations on the highly degenerate classical ground states. The so-called  ``order-by-disorder'' mechanism often comes into play~\cite{Villain, Kawamura, Henley, Bergman}.

\begin{figure}[t]
  \includegraphics[bb=120 0 612 550, width=7.0cm,angle=0]{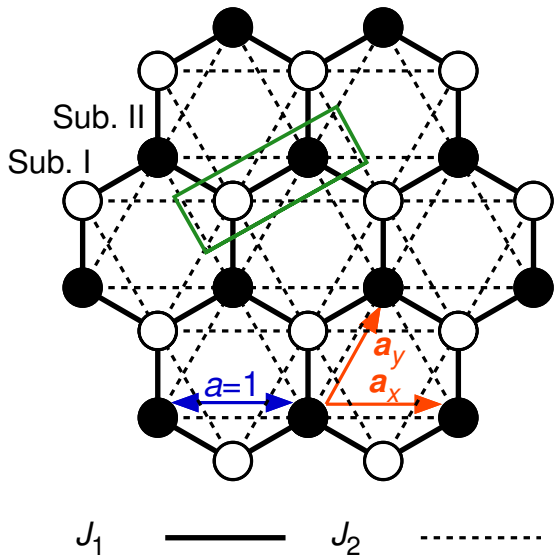}%-90->0
 \caption{(Color online) The $J_1$-$J_2$ honeycomb model with a trigonal symmetry, where $J_1$ and $J_2$ represent the nearest-neighbor (black solid line) and the next-nearest-neighbor (black dot line) interactions. The lattice constant of the triangular lattice $a$, which is equal to the next-nearest-neighbor distance of the honeycomb lattice, is taken to be the length unit, {\it i.e.}, $a=1$. Unit vectors on the triangular lattice are ${\bf a}_x=(a,0)$ and ${\bf a}_y=(\frac{a}{2},\frac{\sqrt{3}a}{2})$. The honeycomb lattice has two lattice sites in a unit cell belonging to two triangular sublattices, which we denote I (white site) and II (black site). \textcolor{black}{Our choice of the unit cell is indicated by the green box. The shape of the honeycomb cluster we mainly treat in this study is a hexagonal one with the trigonal symmetry. The depicted cluster contains 24 spins ($L=4$) under open boundary conditions.}}
 \label{honeycluster}
\end{figure}

 An intriguing example of such frustrated magnets with heavily degenerate classical ground-state manifold might be the antiferromagnetic (AF) Heisenberg model on a honeycomb lattice with the competing nearest-neighbor (NN) $J_1$ and next nearest-neighbor (NNN) $J_2$ couplings as illustrated in Fig.~\ref{honeycluster}.

\begin{figure}[t]
  \includegraphics[bb=130 50 612 452,width=9.4cm,angle=0]{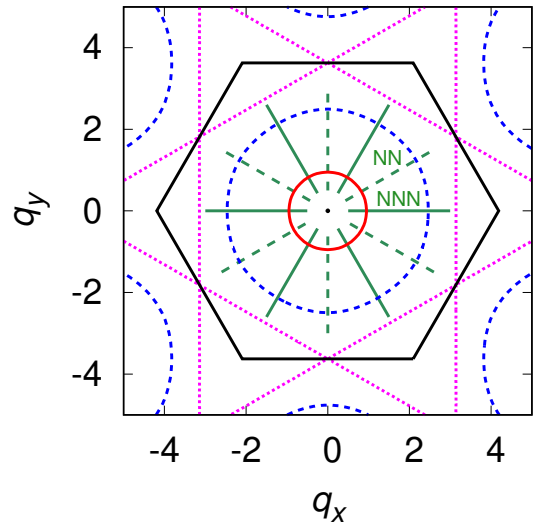}%-90->0
 \caption{(Color online) The ground state manifold of the $J_1$-$J_2$ classical honeycomb-lattice Heisenberg antiferromagnet in the sublattice wavevector space. The red, blue and purple lines represent the degenerate lines for $J_2/J_1=0.18$, $0.3$ and $0.5$, respectively. Black point is located at the origin. Note that the origin corresponds to the \textcolor{black}{wavevector point having a simple two-sublattice Neel order on the original honeycomb lattice.}
 Black solid hexagon indicates the first Brillouin zone of the triangular sublattice. Green dotted and solid lines show the NN and NNN directions, respectively.
}
 \label{gs_generacy}
\end{figure}

 For smaller $J_2/J_1 \leq 1/6$ where the frustration is relatively weak, a simple AF order is stabilized in the ground state by reflecting the bipartite character of the honeycomb lattice. For larger $J_2/J_1 > 1/6$, the ground state becomes a helical or spiral state characterized by a single wavevector ${\bf q}$ which is generally incommensurate with the underlying honeycomb lattice~\cite{Katsura}, a single-$q$ state. An interesting feature here is that the ground-state manifold possesses a macroscopic degeneracy associated with the running directions of the wavevector ${\bf q}$. In the wavevector space, the set of the ground-state $q$'s forms a closed curve surrounding \textcolor{black}{the origin in the sublattice wavevector space.} %AF point.
For $1/6<J_2/J_1<0.5$, this closed curve looks like a ``ring" \textcolor{black}{as demonstrated in Fig.~\ref{gs_generacy}}. This ``ring-like'' degeneracy could give rise to a variety of unique ordering properties~\cite{Okumura, Mulder}. In fact, the effects of thermal and quantum fluctuations in this $J_1$-$J_2$ honeycomb-lattice model have recently been investigated quite intensively~\cite{Okumura,Mulder,Clark, Mosadeq,Cabra,Ganesh,Reuther,Albuquerque,Bishop,Mezzacapo,Bishop2,Zhang,Bishop3,Rosales,Gong,Ganesh2,Zhu,Gong2,Zhang2,Gong3}. 

 One candidate material of the $J_1$-$J_2$ honeycomb-lattice Heisenberg model might be the $S=3/2$ compound ${\rm Bi_{3}Mn_{4}O_{12}(NO_{3})}$~\cite{Smirnova,Matsuda,SOkubo,Azuma, SOkubo2,Onishi}. This compound exhibits a spin-liquid-like behavior without any magnetic long-range order (LRO) down to 0.4 K in spite of a large Weiss temperature of $\sim -257$K. Furthermore, neutron-scattering measurements have revealed that it exhibits a field-induced antiferromagnetism, {\it i.e.\/}, a metamagnetic transition occurs even under relatively weak fields ~\cite{Matsuda}.

 Such unique features of ${\rm Bi_{3}Mn_{4}O_{12}(NO_{3})}$ including the spin-liquid behavior and the field-induced antiferromagnetism were theoretically investigated by Okumura, Kawamura, Okubo and Motome by a Monte Carlo (MC) simulation and a low-temperature expansion~\cite{Okumura}. They have found that the energy-scale of the order-by-disorder is suppressed near the AF phase boundary ($J_2/J_1=$1/6) down to extremely low-temperatures, and the two kinds of exotic spin-liquid states, which are called ``ring-liquid'' and ``pancake-liquid'' states, appear in the low-temperature region. The spin structure factor in the former state exhibits a ring-like pattern surrounding \textcolor{black}{the origin in the sublattice wavevector space}, while, in the spin structure factor in the latter state, the center of the ring-like pattern is ``buried'' in intensity, yielding a ``pancake-like'' pattern. The ring radius of the spin structure factor just corresponds to the radius of the degenerate ring (closed curve) of the ground-state manifold. Okumura {\it et al\/} discussed the possible relationship of these ring-liquid and pancake-liquid states to the experimental properties of ${\rm Bi_{3}Mn_{4}O_{12}(NO_{3})}$, and emphasized the crucial importance of the ring-like degeneracy, which is a source of various exotic spin-liquid-like behaviors and the field-induced antiferromagnetism. At low enough temperatures, the mechanism of order by disorder works, leading to an entropic selection of a particular ${\bf q}$ on the degenerate ring and to a thermodynamic phase transition into the symmetry-broken single-$q$ spiral state.

 From a theoretical viewpoint, an incommensurate ordering as often realized in the frustrated classical $J_1$-$J_2$ model gives rise to a variety of multiple-$q$ states, especially under applied magnetic fields~\cite{Okubo, Okubo2, Rosales, Kamiya, Giacomo, Leonov, ShiZeng, Seabra, Rousoch, Hayami, Janssen}. The multiple-$q$ state is a coherent superposition of states with equivalent but distinct wavevectors related by the underlying symmetry of the lattice.

 For the $J_1$-$J_2$ Heisenberg model on the triangular lattice under magnetic fields, Okubo, Chung and Kawamura identified a variety of multiple-$q$ states including single-$q$, double-$q$ and triple-$q$ states. Especially interesting might be the triple-$q$ state, which corresponds to the {\it skyrmion-lattice\/} state. There, the skyrmion lattice is solely stabilized by the {\it symmetric\/} exchange interaction, and hence, in contrast to the standard skyrmion lattice stabilized by the antisymmetric Dzyaloshinskii-Moriya (DM) interaction, the skyrmion with an opposite sense of the skyrmion number or the spin scalar chirality, {\it i.e.\/}, the antiskyrmion, is also possible. 

 Emergence of such multiple-$q$ states are naturally expected for incommensurate orderings on other lattices as well, {\it e.g.\/}, the honeycomb lattice possessing a common trigonal symmetry with the triangular lattice. On the basis of such an expectation, we study in the present paper the ordering properties of the classical $J_1$-$J_2$ honeycomb-lattice AF Heisenberg model under magnetic fields. As emphasized above, unique feature of the honeycomb-lattice model might be that it exhibits a ring-like continuous degeneracy in its ground state in sharp contrast to the triangular-lattice model, which might give rise to still exotic multiple-$q$ ordered states different from the ones identified in the triangular-lattice model. With this expectation, we study here the $J_1$-$J_2$ honeycomb model in the parameter range of $1/6<J_2/J_1<0.5$ where the ground state of the model exhibits a ring-like infinite degeneracy. \textcolor{black}{As shown in Fig.~2, this ring gets closer to a true circle as $J_2/J_1 \rightarrow 1/6$ while its shape tends to deviate more from a true circle for larger $J_2/J_1$.}   

 Main focus of our simulation is on the case of $J_2/J_1=0.3$, which is located in the middle of the paramagnetic (ring-liquid)-helical phase boundary~\cite{Okumura}. \textcolor{black}{At this value of $J_2/J_1$, the degenerate ring is still close to a true circle (see Fig.~2). Its radius is $q^*\simeq 2.494$ in the NN direction, and is is $q^*\simeq 2.462$ in the NNN direction. When the associated ordered state is to be a single-$q$ spiral state, this $q$-value corresponds to a turn angle on the triangular sublattice of \textcolor{black}{$0.794\pi$ (NN)}, or of \textcolor{black}{$0.784\pi$ (NNN)}, respectively.}

 For this value of $J_2/J_1=0.3$, we indeed find a variety of novel multiple-$q$ ordered states there,  many of which differs in nature from the ones identified in the triangular $J_1$-$J_2$ model. In particular, we observe three different types of double-$q$ state, one of which is essentially of the same nature as the double-$q$ state identified in the triangular model, while the other two are new ones. One is a coplanar state, and the other is a noncoplanar state where the spins form interweaving ``meron-like'' vortex/antivortex lattice pattern. By contrast, only one type of triple-$q$ state is stabilized, which is a collinear state distinct from the skyrmion-lattice state. The triple-$q$ (collinear) state is adiabatically identical with the ``Z state'' identified in the triangular model. Meanwhile, the transverse spin correlation length stays very short in the triple-$q$ (collinear) state of the present model so that the state does not look like the random-domain state consisting of skyrmion and antiskyrmion lattices as observed in the Z phase of the triangular model.

Concerning the single-$q$ states, we find two types,  the umbrella (spiral)-type and the fan-type, the latter stabilized only in higher magnetic fields. For the umbrella-type single-$q$ state, we observe a switching of the running direction of the associated ${\bf q}$ vector with varying the magnetic-field intensity and the temperature. For fuller understanding of the ordering process of the model, we also investigate other $J_2/J_1$-values including $J_2/J_1=0.20, 0.25, 0.35$ and $0.45$.

 The present paper is organized as follows. In Sec.~II, we present our model and explain the numerical and theoretical methods employed. Sec.~III is the main part of the present paper, and is devoted to the presentation of the results of our MC simulations on the model with $J_2/J_1=0.3$.  In \textcolor{black}{Sec.~IV}, we deal with the other $J_2/J_1$ values, $J_2/J_1=0.20, 0.25, 0.35$ and $0.45$. We summarize our main findings in \textcolor{black}{Sec.~V.}  Details of the low-temperature expansion are given in Appendix A, \textcolor{black}{whereas some additional information about the triple-$q$ (collinear, type 2) state stabilized at $J_2/J_1=0.45$  is given in Appendix B}.
\textcolor{black}{In order to get some insights into the relative stability of various multiple-$q$ states, we perform a mean-field (MF) analysis of the model, and compare the results with the MC results. The details are given in the Supplemental Material.}

\begin{figure*}[t]
  \includegraphics[bb=230 60 610 550 ,width=10.0cm,angle=0]{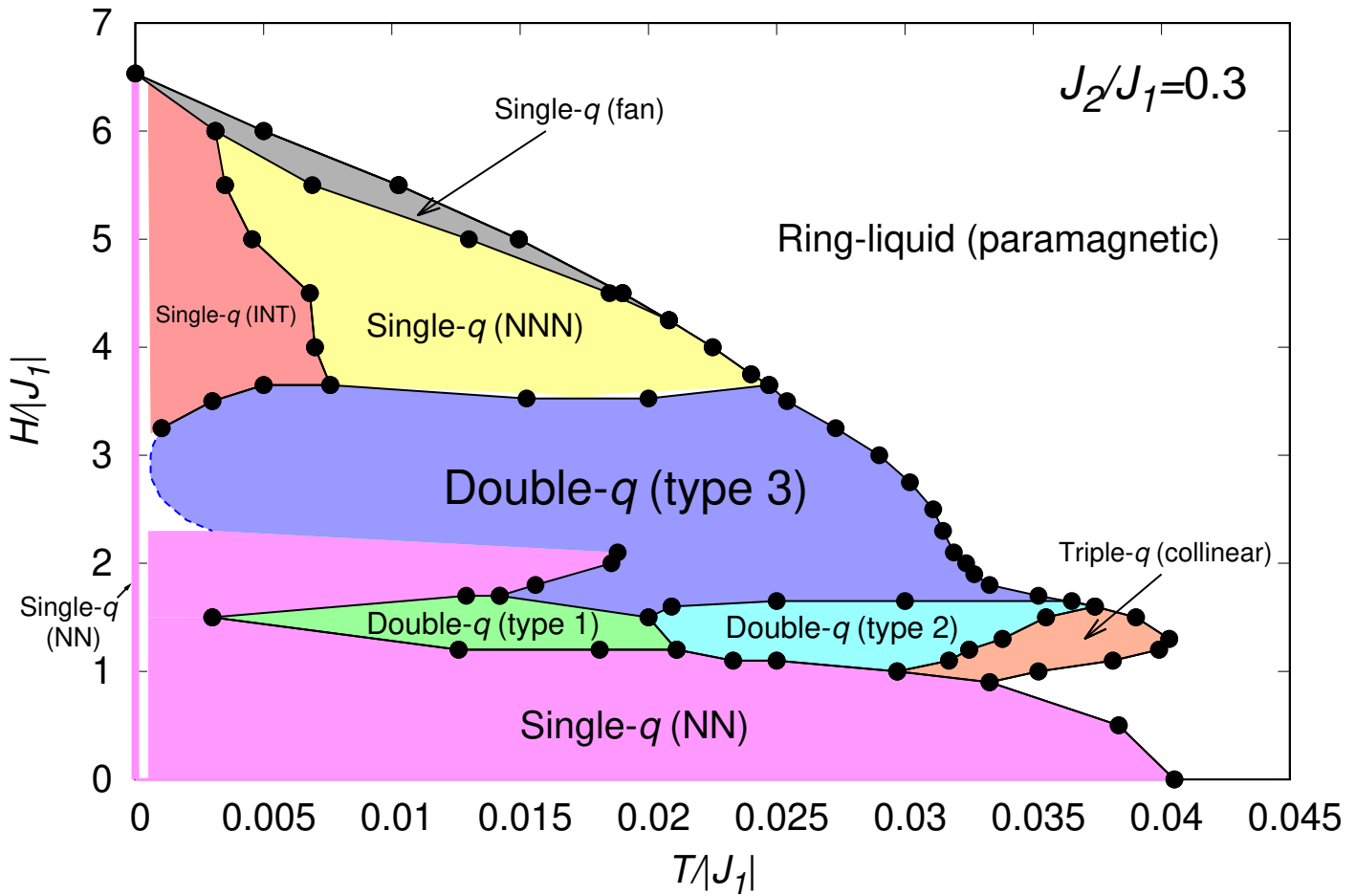}%-90->0
 \caption{(Color online) The $H$-$T$ phase diagram of the $J_1-J_2$ honeycomb-lattice Heisenberg model with $J_2/J_1=0.3$ determined by MC simulations. Transition points are determined from the specific-heat peak, the anomaly in the three-fold symmetry order parameter $m_3$, and that in the magnetic susceptibility (see also Figs.~\ref{temp-dep} and \ref{field-dep}). The dotted blue line representing the low-temperature phase boundary of the double-$q$ (type 3) state remains somewhat arbitrary. The details of each phase are explained in the main text.}
 \label{phase_diagram}
\end{figure*}

\begin{figure}[t]
  \includegraphics[bb=100 20 612 592,width=9.5cm,angle=0]{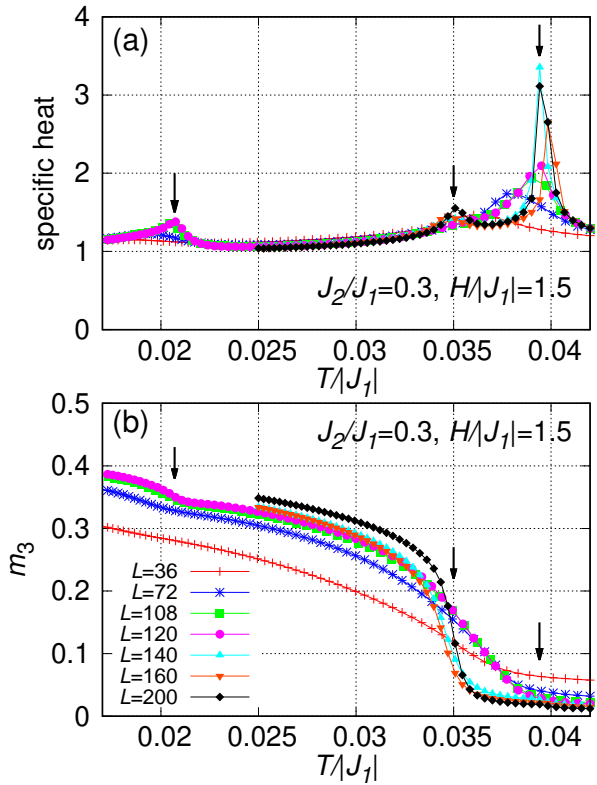}%-90->0
 \caption{(Color online) The temperature dependence of the physical quantities employed in drawing the phase diagram of Fig.~\ref{phase_diagram}, {\it i.e.\/}, (a) the specific heat per spin, and (b) the $m_3$ order parameter. The magnetic-field intensity is $H/|J_1|=1.5$. Arrows indicate the transition points.}
 \label{temp-dep}
\end{figure}

\begin{figure}[t]
  \includegraphics[bb=100 20 612 592, width=9.5cm,angle=0]{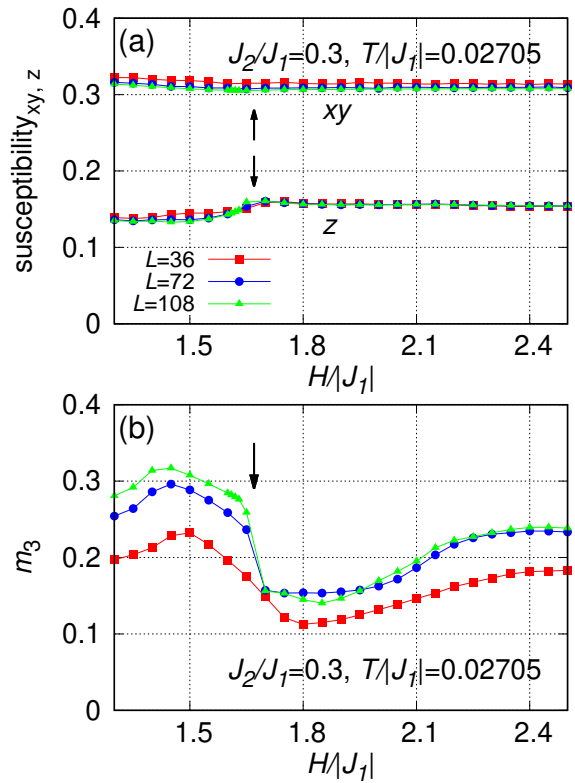}%-90->0
 \caption{(Color online) The field dependence of the physical quantities employed in drawing the phase diagram of $J_2/J_1=0.30$, {\it i.e.\/}, (a) the $xy$- and $z$-components of the magnetic susceptibility, and (b) the $m_3$ order parameter. The temperature is $T/|J_1|=0.02705$. Arrows indicate the transition point.
}
 \label{field-dep}
\end{figure}

\section{\label{sec:Model and Method}II. Model and Method}

We consider the $J_1$-$J_2$ classical honeycomb-lattice Heisenberg model in a magnetic field of intensity $H$, whose Hamiltonian is given by
\begin{eqnarray}
\mathcal{H}=-J_1\sum_{\langle i,j \rangle }{\bf S}_i \cdot {\bf S}_j - J_2\sum_{\langle \langle i,j \rangle \rangle}{\bf S}_i \cdot {\bf S}_j - H \sum_{i} S_i^{z},
\label{hami}
\end{eqnarray}
where ${\bf S}_i=(S_i^x, S_i^y, S_i^z)$ is the classical Heisenberg spin with the fixed length of $|{\bf S}_i|=1$ located at the $i$-th site on the honeycomb lattice, $J_1<0$ and $J_2<0$ represent the antiferromagnetic NN and NNN interactions, while the $\sum_{\langle i,j \rangle}$ and $\sum_{\langle \langle i,j \rangle \rangle}$ are taken over all NN and NNN pairs, respectively. 

It has been known that the ground state of the model in zero field exhibits a single-$q$ helical order for $J_2/J_1>1/6$, with an incommensurate wavevector with an infinite ring-like degeneracy in the $q$-space, while the standard two-sublattice antiferromagnetic order arises for $J_2/J_1\leq 1/6$~\cite{Katsura}.

In general, the multiple-$q$ states are incompatible with the fixed spin-length condition $|{\bf S}_i|$=1 imposed in the ground state, and are not favored in the low temperature region in the classical Heisenberg spin system. Indeed, the multiple-$q$ states have not been reported in previous zero-field calculations of the present model, only a single-$q$ spiral state stabilized with a \textcolor{black}{wavevector selected from the degenerate ring by thermal fluctuations, breaking the threefold discrete $C_3$ lattice symmetry~\cite{Okumura}.} In the present paper, we wish to investigate by means of a MC simulation the possible emergence of the multiple-$q$ states at moderate temperatures under magnetic fields.

MC simulations are performed on the basis of the standard heat-bath method combined with the over-relaxation\textcolor{black}{~\cite{Creutz,Kanki}} and temperature-exchange\textcolor{black}{~\cite{Hukushima}} methods. Our unit MC step consists of one heat-bath sweep and 5-10 over-relaxation sweeps. Typically, our MC runs contain $\sim 10^7$ MC steps, and \textcolor{black}{averages are made over three independent runs in most cases}. In computing certain physical quantities such as the spin structure factor, the temperature-exchange process is stopped to appropriately monitor the symmetry-breaking pattern.

We treat mainly hexagonal finite-size clusters with a trigonal symmetry as illustrated in Fig.~\ref{honeycluster} under open boundary conditions. Due to the enhanced effects of incommensurability, we check the stability of our results also by employing the diamond-shape clusters under periodic boundary conditions. The hexagonal clusters contain $N$=$(3/2)L^2$ spins, where $N$ is the total number of spins on the honeycomb lattice, and we treat the range of sizes $36 \leq L \leq 300$. 

In order to get information about the wavevector of the relevant single-$q$ state at low temperatures, we also employ the low-temperature expansion technique~\cite{Bergman, Okumura}. The details of the calculation are given in Appendix A.
\textcolor{black}{In order to get information about the possible multiple-$q$ ordered states of the model, we also perform the mean-field (MF) analysis. Our MF analysis is the Landau-type free energy expansion up to quartic order following the method of Reimers $et \ al$~\cite{Reimers} and of Okubo $et \ al$~\cite{Okubo,Okubo2}. The details are shown in the Supplemental Material.}

\begin{figure}[t]
  \includegraphics[bb=280 90 542 480,width=3.5cm,angle=0]{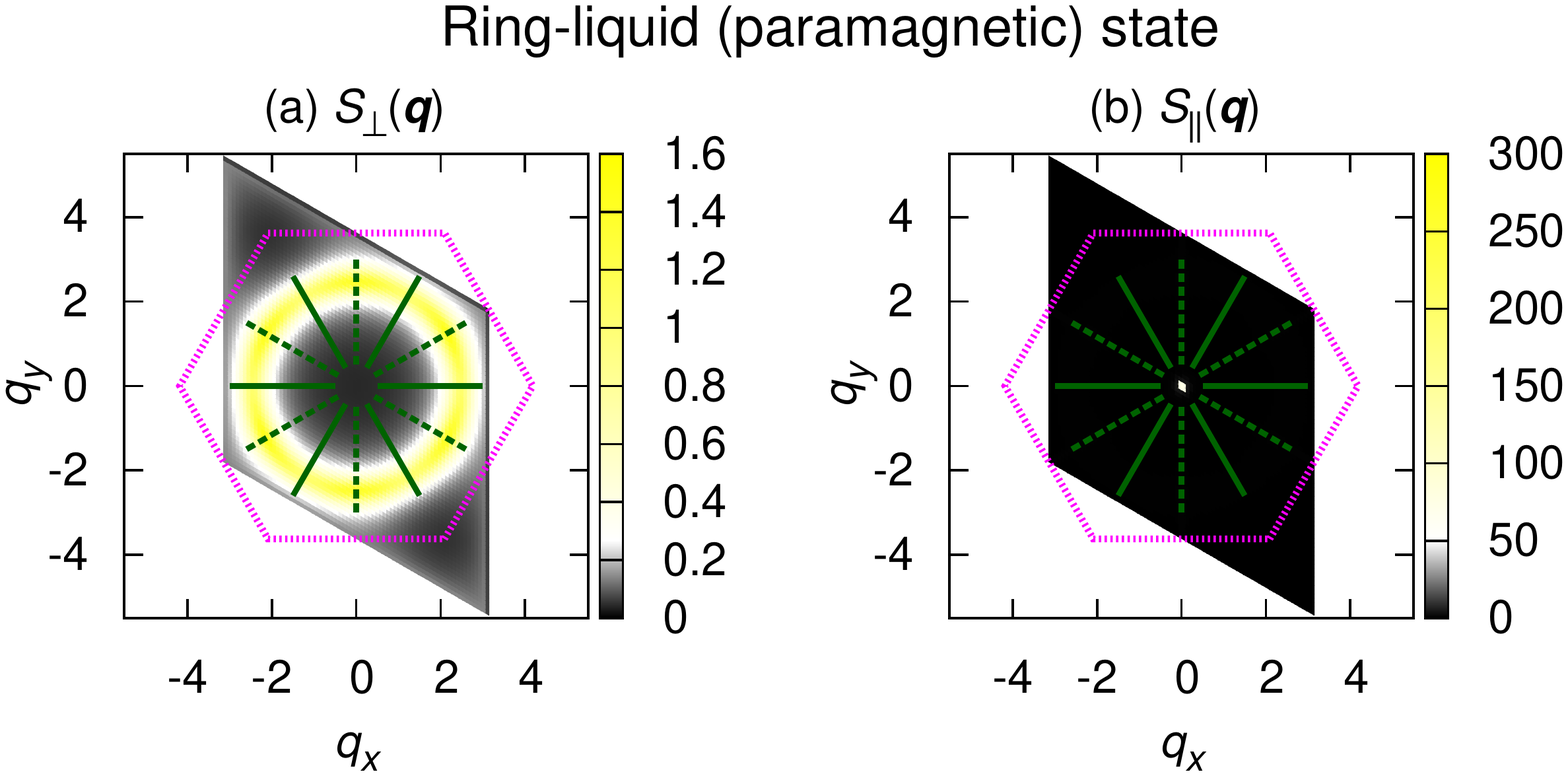}%-90->0
 \caption{(Color online) The intensity plots of the sublattice spin structure factors in the wavevector ($q_x$, $q_y$) plane in the ring-liquid paramagnetic state; (a) the transverse component $S_{\perp}({\bf q})$, and (b) the longitudinal component $S_{\parallel}({\bf q})$. The parameters are $J_2/J_1=0.3$, $H/|J_1|=2.50$ and $T/|J_1|=0.042$, for the lattice size $L=72$. The length unit is taken to be the NNN distance of the honeycomb lattice (or the NN distance of the triangular sublattice). The dotted purple line depicts the zone boundary of the first Brillouin zone of the triangular sublattice. The NN (NNN) directions of the honeycomb lattice are given by the green broken (solid) lines.
 }
 \label{ring_liquid}
\end{figure}

\begin{figure*}
  \includegraphics[bb=180 100 612 492, width=10.5cm,angle=0]{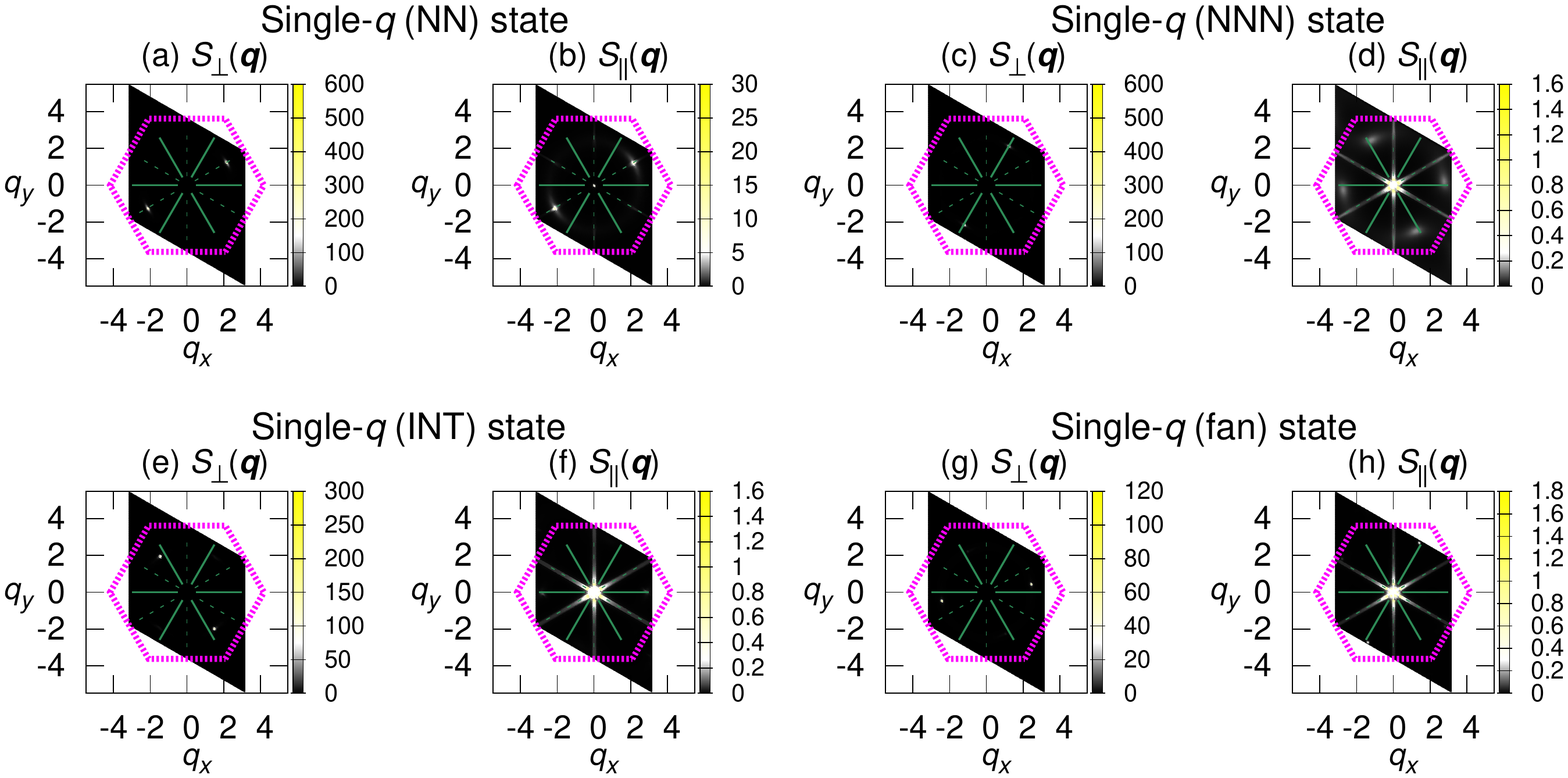}%-90->0
 \caption{(Color online) The intensity plots of the sublattice spin structure factors in the wavevector ($q_x$, $q_y$) plane for various single-$q$ states realized for $J_2/J_1=0.3$. Figs.(a) and (b) correspond to the single-$q$ (NN) state at $H/|J_1|=0.5$, $T/|J_1|=0.0215$ and $L=108$, (c) and (d) to the single-$q$ (NNN) state at $H/|J_1|=4.0$, $T/|J_1|=0.0182$ and $L=120$, (e) and (f) to the single-$q$ (INT) state at $H/|J_1|=5.5$, $T/|J_1|=0.003075$ and $L=120$, and (g) and (h) to the single-$q$ (fan) state at $H/|J_1|=6.0$, $T/|J_1|=0.003862$ and $L=150$. Figs.(a),(c), (e) and (g) represent the the transverse component $S_{\perp}({\bf q})$, while (b),(d), (f) and (h) the longitudinal component $S_{\parallel}({\bf q})$. Note that we tune the intensity range in (d), (f) and (h) \textcolor{black}{to focus subtle features of the intensities on the degenerate ring except for a dominant ${\bf q}={\bf 0}$ peak.} 
}
\label{single_q_matome1}
\end{figure*}

\section{\label{sec:Monte Carlo simulation}III. Monte Carlo results for $J_2/J_1$=0.3}
In this section, we present our MC results. We focus here on the case of $J_2/J_1=0.3$ to study typical ordering patterns arising from the ring-liquid paramagnetic state. As has been demonstrated in the zero-field calculation of Ref.~\cite{Okumura}, the ordered state in zero-field is always a single-$q$ spiral state, not the multiple-$q$ states. The single-$q$ spiral state is generated \textcolor{black}{with a wavevector selected from the degenerate ring-like manifold via the order-by-disorder mechanism, breaking the $C_3$ lattice symmetry of the Hamiltonian.}  For $J_2/J_1=0.3$, the transition temperature is located at $T/|J_1| \simeq 0.0405$~\cite{Okumura}. In this section, we construct a phase diagram in the temperature ($T$) versus magnetic-field ($H$) plane at $J_2/J_1=0.3$.

The obtained $T$-$H$ phase diagram is shown in Fig.~\ref{phase_diagram}. In addition to the single-$q$ states, various types of multiple-$q$ states, including the three distinct types of double-$q$ states and one triple-$q$ state, are also stabilized under magnetic fields due to thermal fluctuations.

In this phase diagram, the transition points are determined mainly from the peak position of the specific heat. As an example, we show in Fig.~\ref{temp-dep}(a) the temperature and size dependence of the specific heat at $H/|J_1|=1.5$. Three sharp peaks appear, each of which corresponds, from high to low temperatures, to the transition from the ring-liquid paramagnetic to the triple-$q$ (collinear) states, to the one from the triple-$q$ (collinear) to the double-$q$ (type 2) states, and to the one from the double-$q$ (type 2) to the double-$q$ (type 1) states. Other quantities such as the $m_3$ order parameter describing the $C_3$ lattice-rotational-symmetry breaking are also employed, $m_3$, which is defined by
\begin{eqnarray}
m_3=\langle |{\bf m}_3| \rangle, \ \  {\bf m}_3=\epsilon_1 {\bf e}_1 + \epsilon_2 {\bf e}_2 + \epsilon_3 {\bf e}_3,
\end{eqnarray}
\textcolor{black}{where ${\bf e}_1=(0,1)$, ${\bf e}_2=(-\sqrt{3}/2,-1/2)$ and ${\bf e}_3=(\sqrt{3}/2,-1/2)$, $\epsilon_{1,2,3}$ are the total NN bond energy normalized per bond along the three NN directions}, respectively, and $\langle \cdots \rangle$ is a thermal average. As can be seen from Fig.~\ref{temp-dep}(b), a transition associated with the $C_3$ symmetry breaking is expected at $T/|J_1| \simeq 0.035$, which coincides with the second peak of the specific heat in Fig.~\ref{temp-dep}(a). Although a sharp diverging peak of the specific heat, possibly corresponding to a first-order phase transition, is observed at a higher temperature $T/|J_1| \simeq 0.039$, the $m_3$ order parameter in the thermodynamic limit still remains to be vanishing there, meaning that the transition at $T/|J_1| \simeq 0.039$ is the one keeping the $C_3$ symmetry, {\it i.e.\/}, the transition into the triple-$q$ (collinear) state in the phase diagram of Fig.~\ref{phase_diagram}.

Sometimes, the phase boundary happens to be almost \textcolor{black}{temperature-independent}, {\it i.e.\/}, almost horizontal in the $T$-$H$ phase diagram. In such a case, the magnetic field dependence of physical quantities could also be useful in determining the phase boundary. As an example, we show in Figs.~\ref{field-dep}(a) and (b) the magnetic-field dependence of (a) the $xy$- and $z$- components of the differential magnetic susceptibility, and (b) the $m_3$ order parameter, which turn out to be useful in locating the phase boundary between the double-$q$ (type 2) and (type 3) states as indicated by the arrow in the figure.

 A convenient quantity in identifying various types of multiple-$q$ ordered states might be the  static spin structure factor. In the present paper, in view of the basic two-sublattice (I or II) nature of the ordering, we compute primarily the {\it sublattice\/} spin structure factor, both perpendicular to the field (the $xy$-component) and parallel with the field (the $z$-component). Note that the honeycomb lattice contains two lattice points in its unit cell, each forming a triangular sublattice whose unit lattice vector corresponds to the NNN direction of the original honeycomb lattice: see Fig.~\ref{honeycluster}.

 The $xy$ component of the sublattice spin structure factor $S_{\perp}({\bf q})$ is defined by
\begin{eqnarray}
S_{\perp}({\bf q})=\frac{2}{N}  \sum_{\mu=x,y}\langle |\sum_{j \in {\rm I \ or \ II}} S_j^{\mu} e^{-i {\bf q} \cdot {\bf r}_j}|^2 \rangle,
\end{eqnarray}
while the $z$ component $S_{\parallel}({\bf q})$ by
\begin{eqnarray}
S_{\parallel}({\bf q})=\frac{2}{N}  \langle |\sum_{j \in {\rm I \ or \ II}} S_j^z e^{-i {\bf q} \cdot {\bf r}_j}|^2 \rangle,
\end{eqnarray}
where ${\bf r}_j$ is the position vector of the spin at the $j$-th site on each triangular sublattice, and ${\bf q}=(q_x,q_y)$ is the associated wavevector. Thus, in our present definition of the $q$-vector, \textcolor{black}{the ${\bf q}=(0, 0)$ point corresponds to the wavevector point associated with the two-sublattice Neel order on the original honeycomb lattice.}
\textcolor{black}{In any finite-size simulation, fully symmetric patterns should be obtained in the spin structure factor when the system is fully thermalized, whereas, in the ordered state, such a time scale usually becomes extremely long for a moderately large system. Hence, in computing the spin structure factor in our present MC simulation, we turn off the temperature-exchange process, and monitor the symmetry-breaking pattern typically during $10^{3} \sim 10^4$ MC steps for the measurements.}

\subsection{A. The ring-liquid (paramagnetic) states}

 In Figs.~\ref{ring_liquid}, we show the intensity plots of the sublattice static spin structure factor in the ring-liquid paramagnetic state. As can be seen from the figure, $S_{\perp}({\bf q})$ exhibits a broad ring-like intensity, while $S_{\parallel}({\bf q})$ exhibits a sharp peak only at $q=0$ arising from the uniform magnetization induced by an applied magnetic field. This ring-like intensity reflects the ring-like degeneracy of the ground state as argued above. On decreasing the temperature, various types of ordered states including multiple-$q$ states could emerge \textcolor{black}{by selecting various wavevectors from the ring-like degenerate manifold.}
In this sense, the ring serves as a source of various multiple-$q$ states to be discussed below.

\subsection{\label{sec:The single-$q$ states}B. The single-$q$ states}

The single-$q$ state is characterized by one of the incommensurate wavevectors on the ring, ${\bf q}^{*}$, and its partner $-{\bf q}^{*}$. In fact, there exist several different types of single-$q$ states under magnetic fields at \textcolor{black}{$J_2/J_1=0.3$}, such as the single-$q$ (NN) state, the single-$q$ (NNN) state, the single-$q$ (INT) state, and the single-$q$ (fan) state. \textcolor{black}{The former two states were already referred to in Introduction and reported in ref.~\cite{Okumura}. The single-$q$ (NN, NNN and INT) states have umbrella-type spin textures, while the single-$q$ (fan) state has a different type of spin texture, a fan-like coplanar structure, as will be explained below.}

In the single-$q$ (NN) state, the spiral axis runs along the NN direction of the honeycomb lattice. As can be seen from the phase diagram of Fig.~\ref{phase_diagram}, this state is realized at relatively low magnetic fields including zero magnetic field. In fact, this observation is fully consistent with the previous finding of Okumura {\it et al\/} that the spiral runs along the NN direction in zero field for $J_2/J_1=0.3$~\cite{Okumura}.

 We show in Figs.~\ref{single_q_matome1}(a) and (b) the typical sublattice spin structure factors for the single-$q$ (NN) state. Sharp peaks appear in the transverse component $S_\perp({\bf q})$ at a pair of $\pm {\bf q}^{*}$, while broader peaks appear in the longitudinal component ${S_\parallel({\bf q})}$ at the same wavevector points $\pm {\bf q}^{*}$, in addition to the uniform component at  ${\bf q}={\bf 0}$ induced by an applied field. The broad peaks of ${S_\parallel({\bf q})}$ do not sharpen with increasing the system size, indicating  the short-ranged-order (SRO) character of the $z$ component. In contrast, the sharp feature of  $S_\perp({\bf q})$ is consistent with the expected quasi-LRO character of the single-$q$ spiral structure.

 Stronger magnetic fields can produce a single-$q$ spiral running along the NNN directions, which we denote as a single-$q$ (NNN) state as shown in Fig.~\ref{phase_diagram}. The corresponding spin structure factors are shown in Figs.~\ref{single_q_matome1}(c) and (d). \textcolor{black}{In contrast to sharp peaks of \textcolor{black}{${S_\perp({\bf q})}$}, weak broad peaks appear in \textcolor{black}{${S_\parallel({\bf q})}$} at the wavevectors complementary to the strong spots in \textcolor{black}{${S_\perp({\bf q})}$}.}

 We note that, in the lower temperature region and in stronger magnetic fields, the single-$q$ spiral runs along an intermediate direction between the NN and the NNN ones, which we denote as a single-$q$ (INT) state. The corresponding spin structure factor is shown in Figs.~\ref{single_q_matome1}(e) and (f).

\begin{figure}[t]
  \includegraphics[bb=230 50 610 500, width=4.2cm,angle=0]{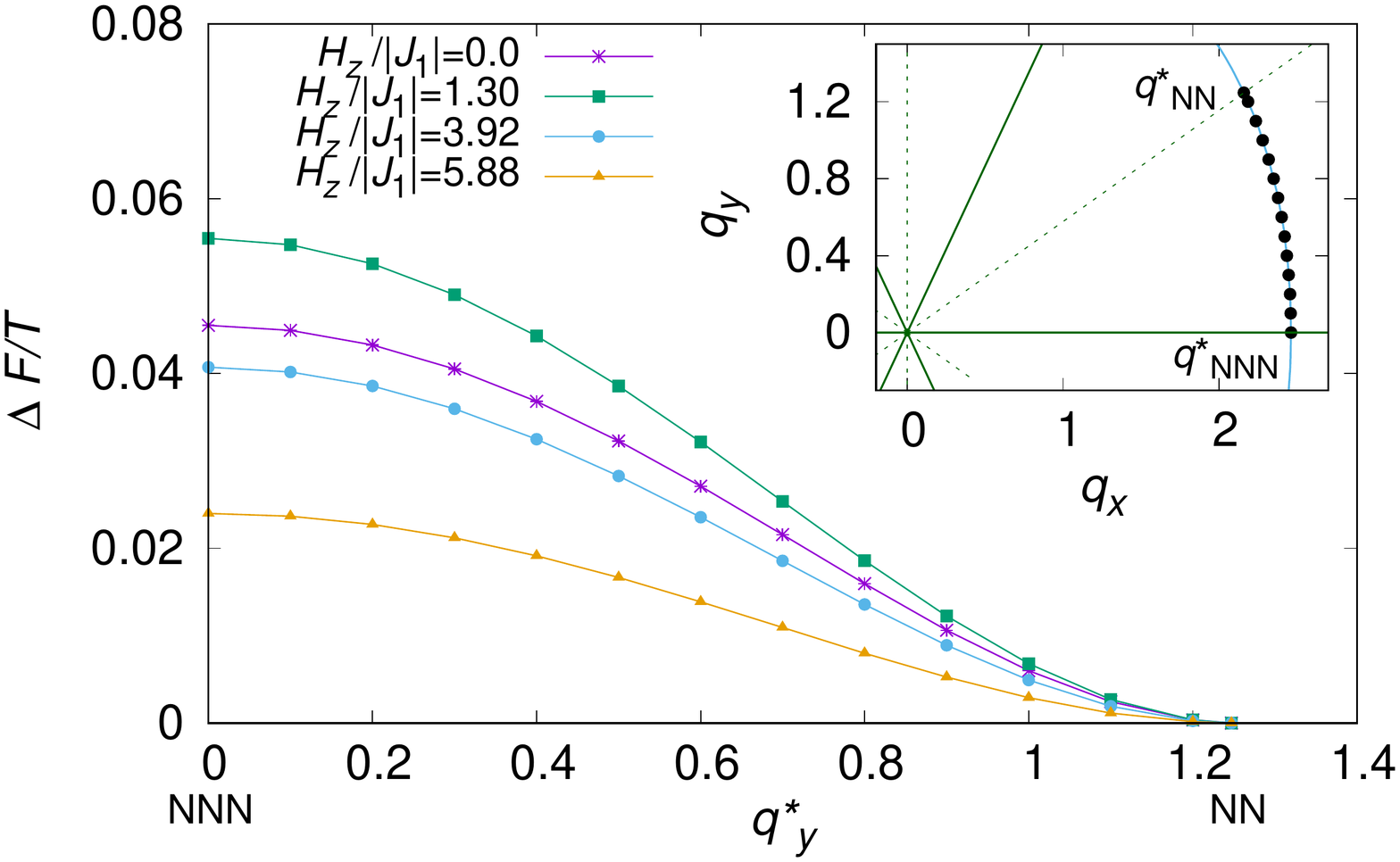}%-90->0
 \caption{(Color online) The $q^*$-vector direction dependence of the free-energy density difference ${\Delta F}/T \equiv F({\bf q}^{*})/T - F({\bf q}_{\rm NN}^{*})/T$ for various fields calculated by the low-temperature expansion at $J_2/J_1=0.30$, where the free energy density of the spiral running along the NN direction $F({\bf q}^{*}_{\rm NN})$ is taken as an energy origin, and ${\bf q}^{*}=(q_x^{*}, q_y^{*})$ is the wavevector on the degenerate ring (the blue line in the inset) for $J_2/J_1=0.3$.  As depicted in the inset, the direction of the spiral is represented by its $q_y^{*}$ value. The NN and the NNN directions are drawn by the green solid and broken lines, respectively.
}
\label{low_temp}
\end{figure}

\begin{figure}[t]
\includegraphics[bb=50 250 580 792, width=10cm,angle=0]{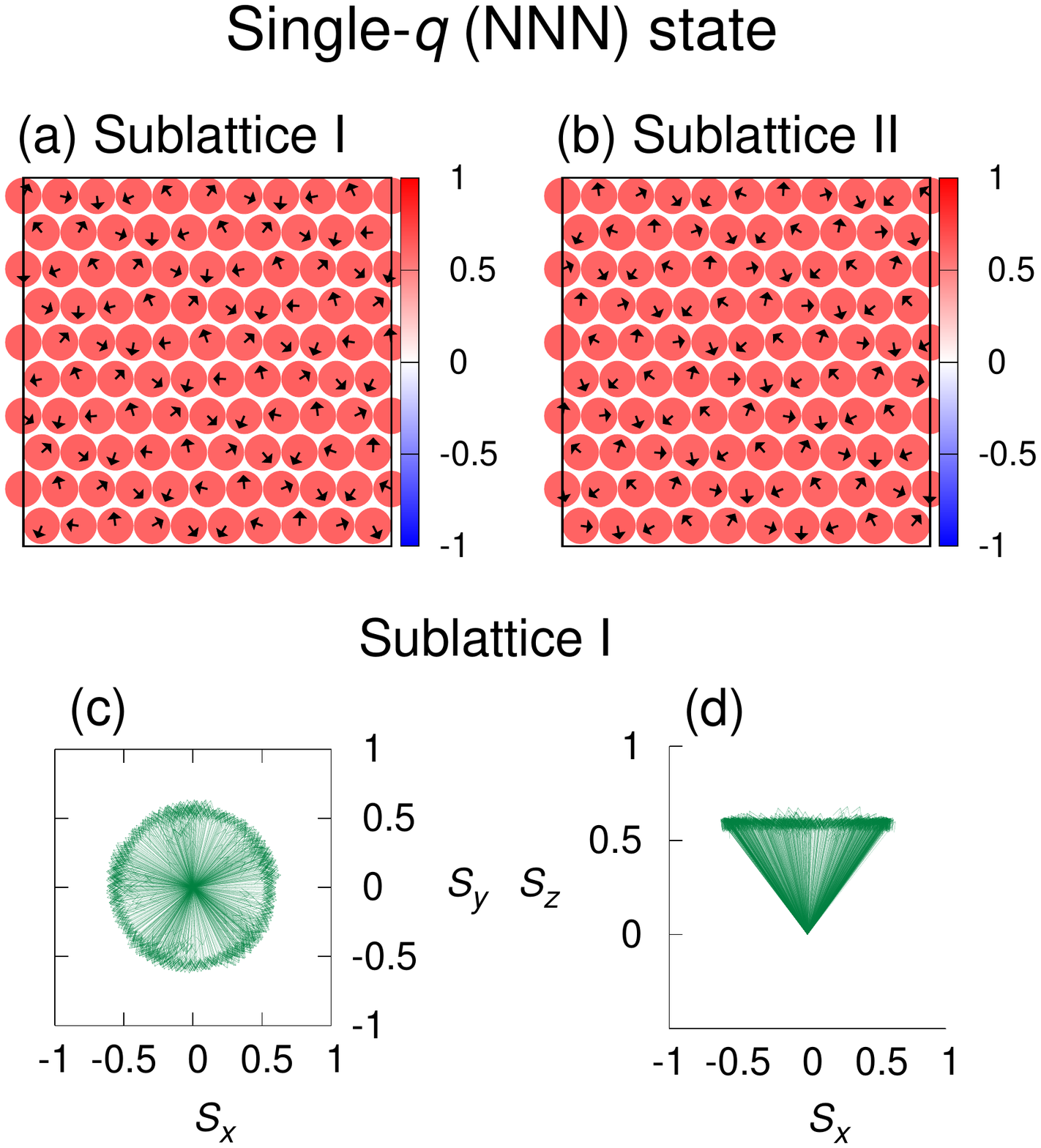}
\caption{(Color online) Real-space sublattice spin configurations in the single-$q$ (NNN) state for $J_2/J_1=0.3$ obtained by the short-time average of 1000 MC steps, \textcolor{black}{(a) for the sublattice I, and (b) for the sublattice II.} The parameters are $H/|J_1|=4.0$, $T/|J_1|=0.01820$ and $L=120$. \textcolor{black}{(a,b)} The $xy$-components of the spin are represented by the arrow, while the $z$ component is represented by the blue-to-red color scale. In \textcolor{black}{(c) and (d), spins at various sites on the sublattice I are reorganized with a common origin, (c) the top view in the ($S_x$, $S_y$) plane, and (d) the side view in the ($S_x$, $S_z$) plane.}
}
\label{single_umbrella}
\end{figure}

\begin{figure}[t]
\includegraphics[bb=50 0 540 792, width=7.3cm,angle=0]{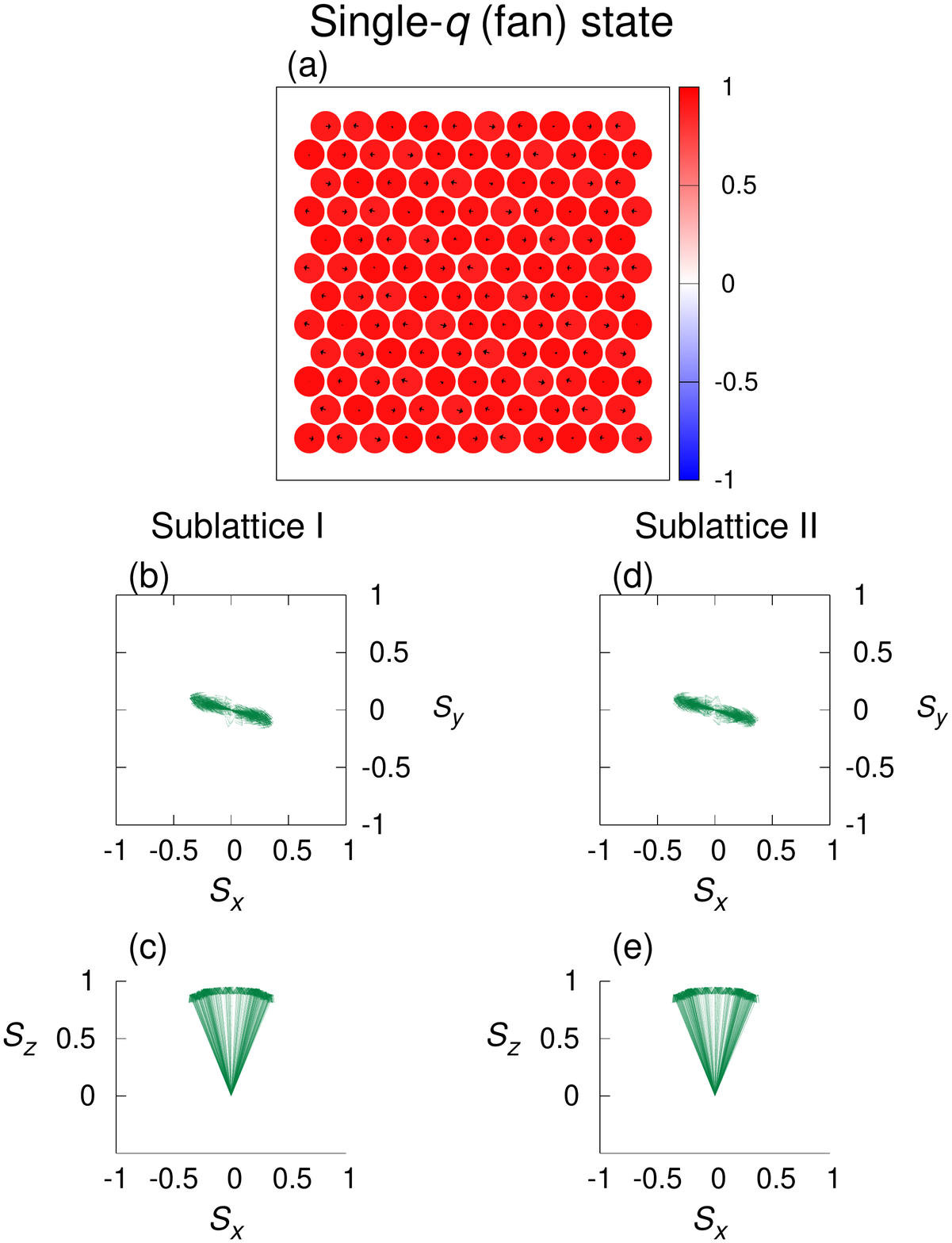}
\caption{(Color online) Real-space sublattice spin configurations in the single-$q$ (fan) state for $J_2/J_1=0.3$ obtained by the short-time average of 1000 MC steps, \textcolor{black}{(a)-(c) for the sublattice I, and (d)(e) for the sublattice II. The parameters are $H/|J_1|=6.0$, $T/|J_1|=0.003862$ and $L=150$. (a) The $xy$-components of the spin are represented by the arrow, while the $z$ component is represented by the blue-to-red color scale. In (b)-(e), spins at various sites on a given sublattice are reorganized with a common origin, (b)(d) the top view in the ($S_x$, $S_y$) plane, and (c)(e) the side view in the ($S_x$, $S_z$) plane.}
}
\label{single_fan}
\end{figure}

Thus, depending on $T$ and $H$, the single-$q$ spiral can run in various directions, while any direction can be selected from the $q$-directions on the degenerate ring. In order to understand the selection mechanism of a particular $q$-direction, {\it i.e.\/}, the ``order-by-disorder'' mechanism operating here, we employ the low-temperature expansion calculations of Refs. \cite{Bergman} and \cite{Okumura} by extending their zero-field calculations to nonzero fields. \textcolor{black}{Note that  the low-temperature expansion is completely independent of our MC calculation, no input provided from MC. Some of the details are explained in Appendix A.} We find that, at $J_2/J_1=0.30$, thermal fluctuations always select the NN direction at low enough temperatures for any $H$. In Fig.~\ref{low_temp}, we show  for $J_2/J_1=0.30$ the directional dependence of the free-energy density difference between a given spiral state with the wavevector ${\bf q}^{*}$ lying on the degenerate ring (see black points in the inset of Fig.~\ref{low_temp}) and the spiral running along the NN direction, \textcolor{black}{$\Delta F/T \equiv F({\bf q}^{*})/T - F({\bf q}_{\rm NN}^{*})/T$}, computed based on eqs.~(\ref{lowT_F})-(\ref{lowT_I_II}). As can be seen from the figure, $\Delta F$ is always positive, indicating that thermal fluctuations prefer the NN direction in the low-temperature limit $T/|J_1| \rightarrow 0$ among all possible directions on the degenerate ring. 

\textcolor{black}{At higher magnetic fields of $H/|J_1|>2.2$, the low-temperature expansion results might seem inconsistent with our MC phase diagram of Fig.~3. One possible cause of this apparent discrepancy might be a possible failure of the harmonic (Gaussian) approximation of our low-temperature expansion neglecting the nonlinear effects. Meanwhile, our MC specific-heat data at $H/|J_1|>2.2$ take values greater than the harmonic value of unity expected for the classical Heisenberg spin systems, down to a low temperature $T/|J_1|=0.005$. In contrast, the specific heat should take a value less than unity~\textcolor{black}{\cite{Zhitomirsky}} when the nonlinear effects around the ground state are dominant. Hence, a plausible explanation here might be that an additional phase transition, most probably to the single-$q$ (NN) state, occurs at a still lower temperature of $T/|J_1|<0.005$. Unfortunately, we could not thermalize such a low-temperature regime in our MC.}

 These three types of single-$q$ states (NN, NNN and INT) have umbrella-type structures in their real-space spin configurations. We show in Fig.~\ref{single_umbrella} typical real-space sublattice spin configurations of the single-$q$ (NNN) state, \textcolor{black}{ (a) for the sublattice I, and (b)  for the sublattice II. These real-space spin configurations correspond to the spin structure factors shown in Figs.~7(c) and (d). In \textcolor{black}{Figs.~\ref{single_umbrella}}(c) and (d), spins at various sites on the sublattice I are reorganized in the spin space with a common origin, a top view in the ($S_x$, $S_y$) plane in (c), and a side view in the ($S_x$, $S_z$) plane in (d). Essentially the same plots are obtained also for the sublattice II (not shown here). The real-space $xy$-spin configurations on the two sublattices look essentially similar, with a phase difference of $\alpha \sim 0.55\pi$. This value of the phase difference is a bit smaller than, but close to the corresponding value expected in the ground state for $J_2/J_1=0.3$, $\alpha \sim 0.61\pi$: Refer to $\alpha_{q^*_{1,{\rm NNN}}}$ in \textcolor{black}{Fig.~S1(a) of the Supplemental Material.} The observed small deviation is most probably the temperature effect.}

 In addition to such umbrella-type single-$q$ states, the other type of single-$q$ state also appears in the high-field region of the phase diagram just below the phase boundary to the ring-liquid paramagnetic state. In this state, as can be seen from Figs.~7(g) and (h), the associated ordering wavevector runs along an intermediate direction between the NN and the NNN directions like the single-$q$ (INT) state. The real-space spin configuration has a fan-like coplanar structure, instead of the noncoplanar umbrella-type one.  We call this type of single-$q$ state a single-$q$ (fan) state. Fig.~\ref{single_fan} exhibits the typical spin configuration of this single-$q$ (fan) state realized at $H/|J_1|=6.00$ and $T/|J_1|=0.003862$ \textcolor{black}{for the sublattice I in (a)-(c), and for the sublattice II in (d) and (e). As can be seen from the figures, the spin structure in the fan state is coplanar.}

\begin{figure}[t]
  \includegraphics[bb=60 0 612 810, width=9.0cm,angle=0]{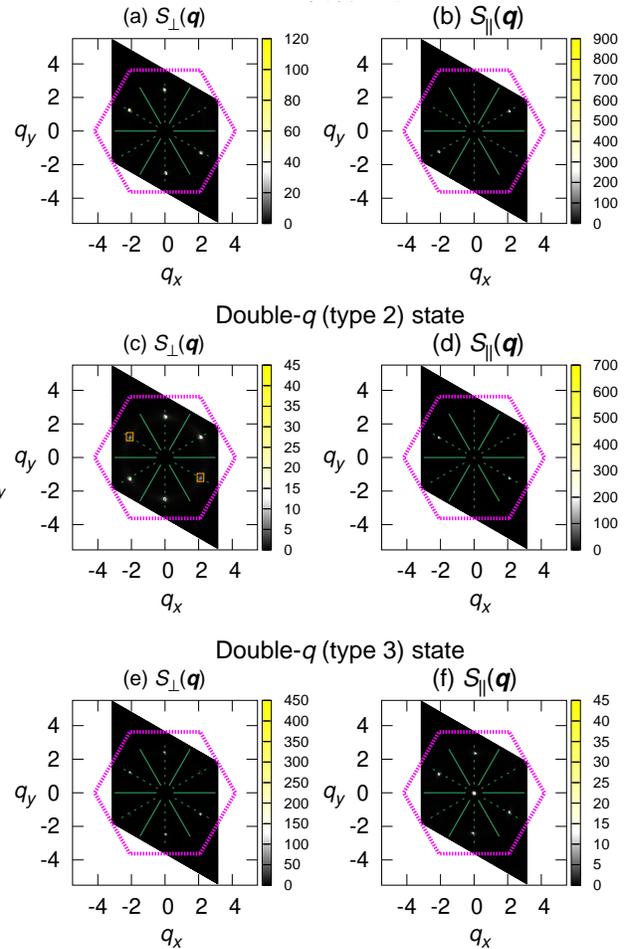}
 \caption{(Color online) The intensity plots of the sublattice spin structure factors in the wavevector ($q_x$, $q_y$) plane in various type of double-$q$ states at $J_2/J_1=0.3$; (a)(c)(e) the transverse component $S_{\perp}({\bf q})$, and (b)(d)(f) the longitudinal component $S_{\parallel}({\bf q})$. The parameters are $H/|J_1|=1.5$, $T/|J_1|=0.017$ and $L=108$ for (a) and (b), $H/|J_1|=1.5$, $T/|J_1|=0.025$ and $L=108$ for (c) and (d), and $H/|J_1|=2.5$, $T/|J_1|=0.027$ and $L=150$ for (e) and (f).
}
 \label{double-q-matome}
\end{figure}

\begin{figure}[t]
  \includegraphics[bb=50 220 612 792,width=10cm,angle=0]{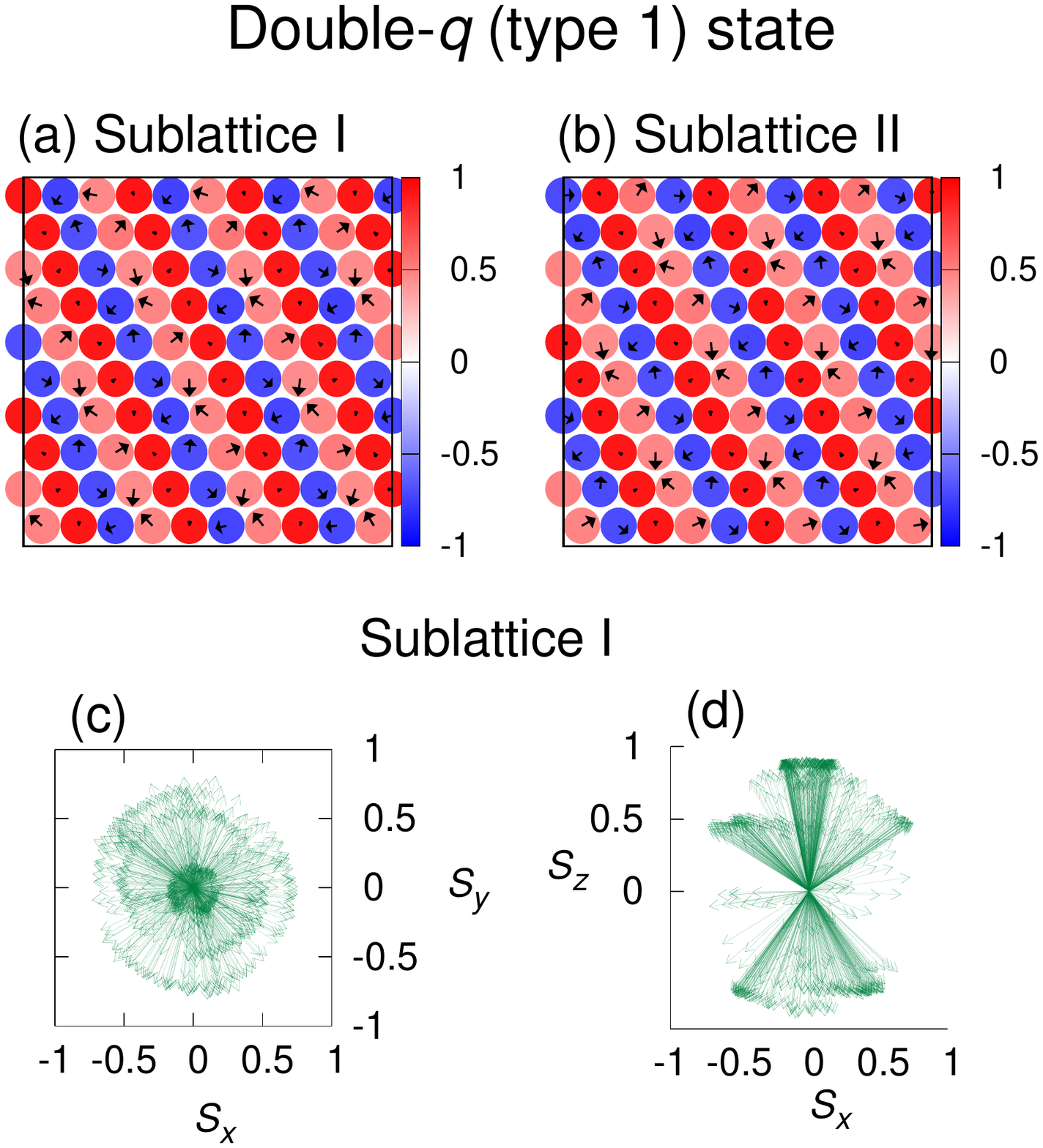}
 \caption{(Color online) Real-space sublattice spin configurations in the double-$q$ (type 1) state obtained by short-time average of 400 MC steps, (a) for the sublattice I, and (b) for the sublattice II. The parameters are $J_2/J_1=0.3$, $H/|J_1|=1.5$, $T/|J_1|=0.017$ and $L=108$. (a,b) The $xy$ components of the spin are represented by the arrow, while the $z$ component is represented by the blue-to-red color scale. In (c) and (d), spins at various sites on the sublattice I are reorganized with a common origin, (c) the top view in the ($S_x$, $S_y$) plane, and (d) the side view in the ($S_x$, $S_z$) plane.
}
 \label{double_type1_snap}
\end{figure}

\begin{figure}[t]
  \includegraphics[bb=0 0 612 792, width=10cm,angle=0]{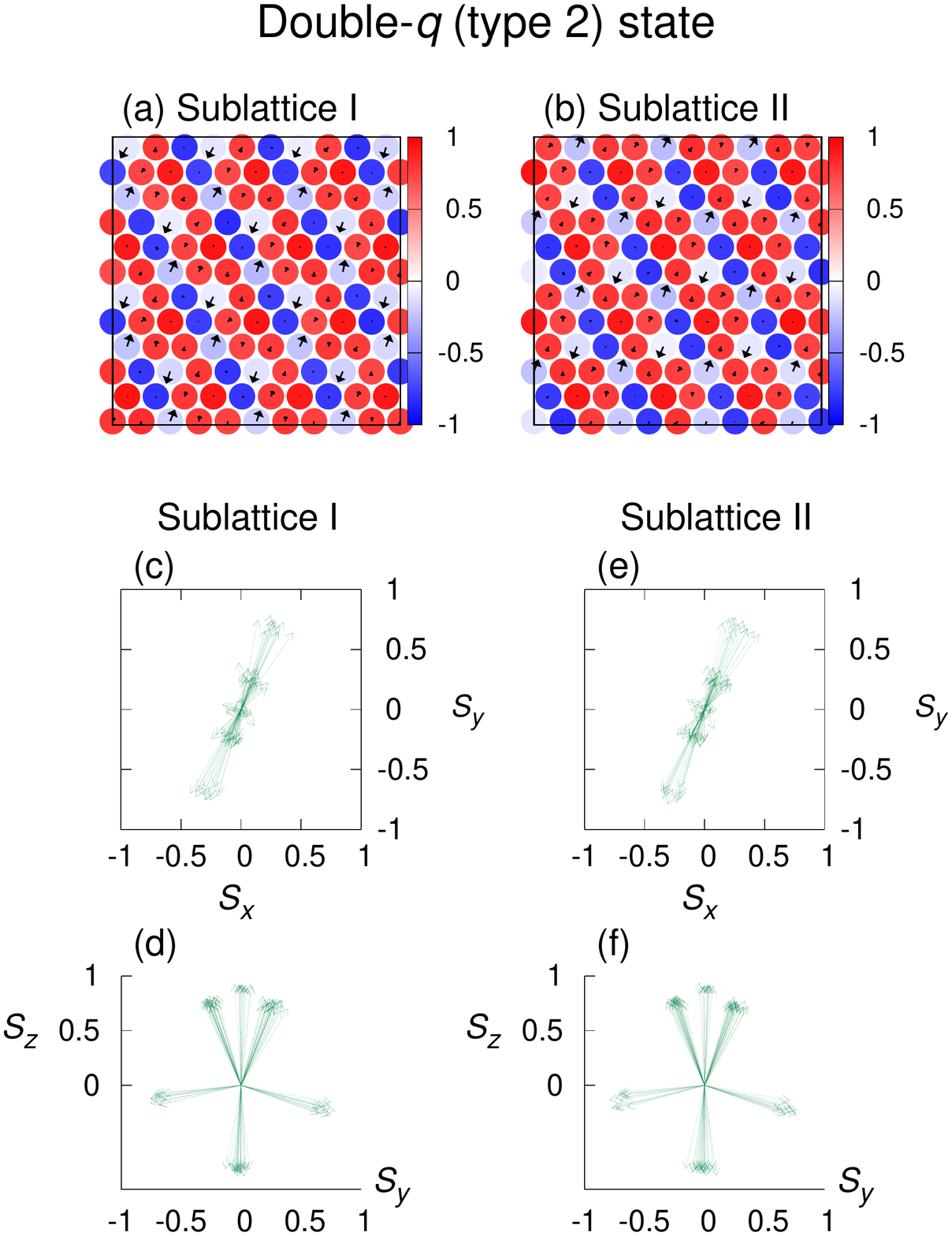}
 \caption{(Color online) Real-space sublattice spin configurations in the double-$q$ (type 2) state obtained by the short-time average of 100 MC steps, (a) for the sublattice I, and (b) for the sublattice II. The parameters are $J_2/J_1=0.3$, $H/|J_1|=1.5$, $T/|J_1|=0.025$ and $L=200$. (a,b) The $xy$ components of the spin are represented by the arrow, while the $z$ component is represented by the blue-to-red color scale. \textcolor{black}{In (c)-(f), spins at various sites on the sublattice I (c,d) and II (e,f) are reorganized with a common origin, (c)(e) the top view in the ($S_x$, $S_y$) plane, and (d)(f) the side view in the ($S_y$, $S_z$) plane.}
}
 \label{double_type2_snap}
\end{figure}

\begin{figure}[t]
 \includegraphics[bb=230 0 612 592, width=5.5cm,angle=0]{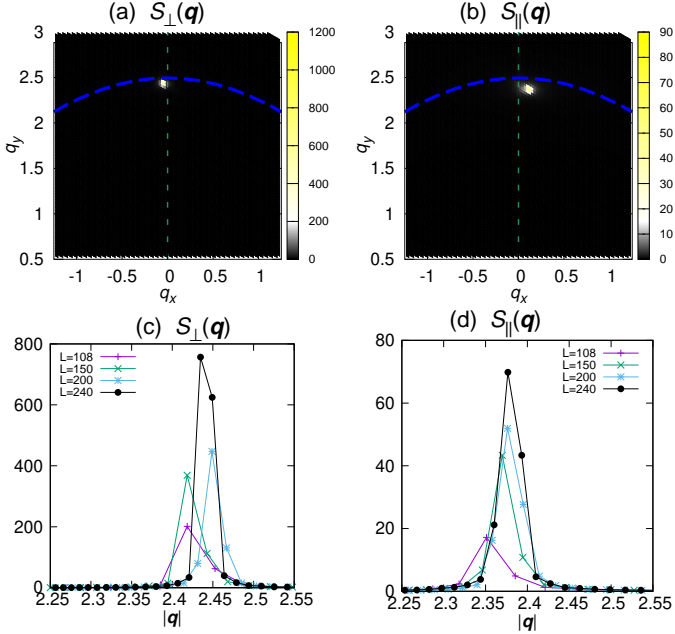}%-90->0
 \caption{(Color online) The intensity plots of the sublattice spin structure factors in the double-$q$ (type 3) state focused around one spot located close to the $q_y$ axis, for (a) the transverse component $S_{\perp}({\bf q})$, and (b) for the longitudinal component $S_{\parallel}({\bf q})$. The parameters are $J_2/J_1=0.3$, $H/|J_1|=2.5$, $T/|J_1|=0.027$ and $L=240$. The blue dotted curve represents the degenerate ring corresponding to $J_2/J_1=0.3$ as shown in Fig.~\ref{gs_generacy}. The wavevector $|{\bf q}|$ dependence of the static spin structure factor peaks of (c) the transverse component $S_{\perp}({\bf q})$, and (d) the longitudinal component $S_{\parallel}({\bf q})$, measured along the line passing the peak position in the NN direction. The parameters are $J_2/J_1=0.3$, $H/|J_1|=2.5$ and $T/|J_1|=0.02708$ for the lattice sizes $L=108, 150, 200$ and $240$. 
}
 \label{double_type3_Sq_size_dep}
\end{figure}

\begin{figure}[t]
  \includegraphics[bb=50 0 612 792, width=9.5cm,angle=0]{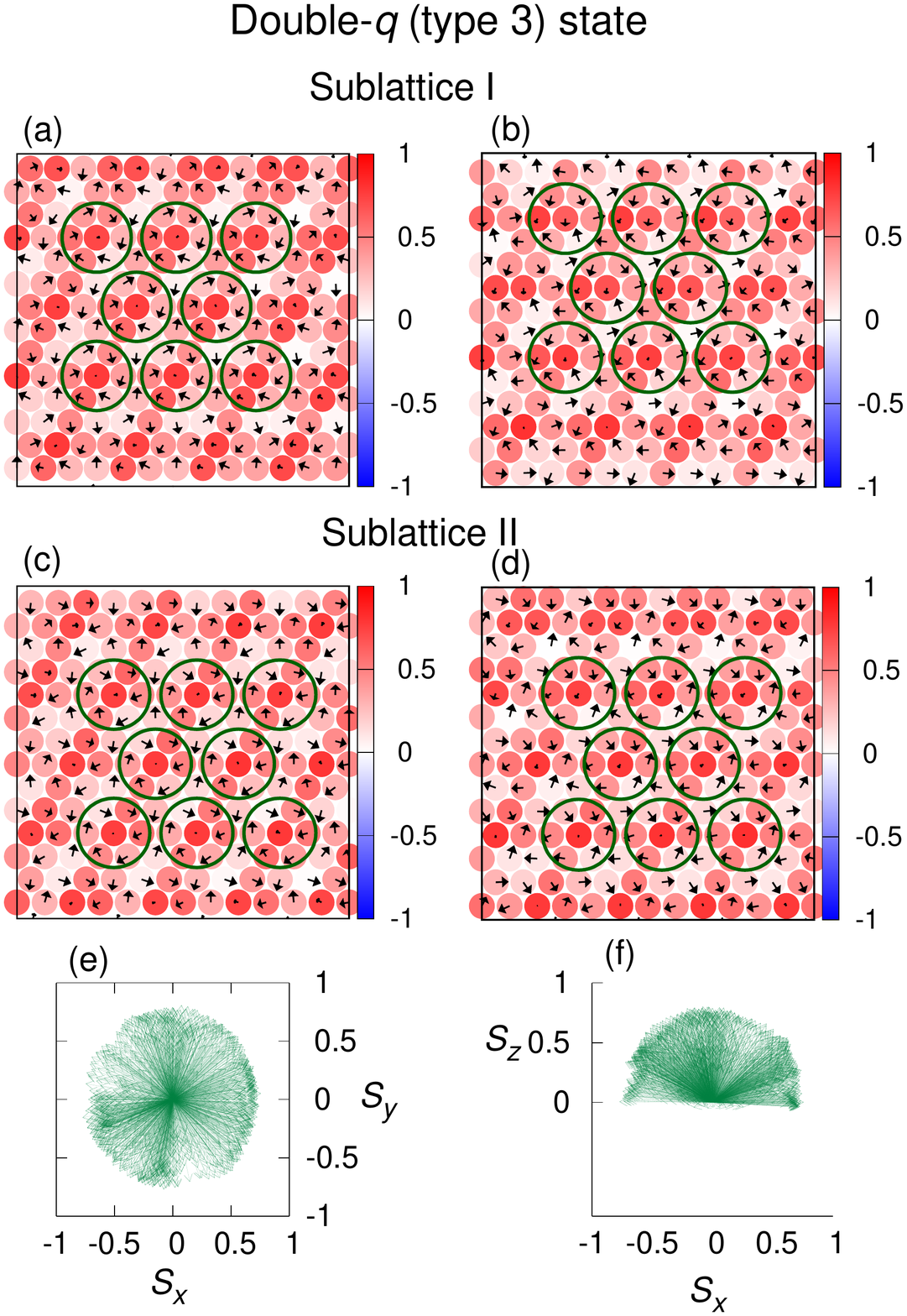}
 \caption{(Color online) (a,b) The real-space sublattice spin configurations of the double-$q$ (type 3) state obtained by the short-time average of 1000 MC steps, (a,b) for the sublattice I, and (c-f) for the sublattice II. The parameters are $J_2/J_1=0.3$, $H/|J_1|=2.5$, $T/|J_1|=0.027$ and $L=240$. The \textcolor{black}{spin configurations} are shown in (a,c) and in (b,d), each set representing two different spatial regions in the same sample. The $xy$ components of the spin are represented by the arrow, while the $z$ component is represented by the blue-to-red color scale. In (e) and (f), spins at various sites on the sublattice II are reorganized with a common origin, (e) the top view in the ($S_x$, $S_y$) plane, and (f) the side view in the ($S_x$, $S_z$) plane.
}
 \label{double_type3_snap}
\end{figure}

\subsection{\label{sec:The double-$q$ (type 1) state}C. The double-$q$ (type 1) state}

The double-$q$ structure is characterized by two pairs of ${\bf q}^{*}$ in $S_\perp({\bf q})$. For $J_2/J_1=0.3$, three distinct types of double-$q$ states appear. Let us begin with the double-$q$ (type 1) state, whose sublattice spin structure factors are shown in Figs.~\ref{double-q-matome}(a) and (b). Two pairs of spot intensities appear in $S_{\perp}({\bf q})$, while a pair of intensities appears in $S_{\parallel}({\bf q})$ at the wavevectors complementary to the two pairs in $S_{\perp}({\bf q})$. In this state, relevant ${\bf q}^{*}$-vectors run along the NN direction. The reason why the ${\bf q}^{*}$ vectors run along the NN directions will be discussed within the MF analysis \textcolor{black}{(see the Supplemental Material)}. In fact, this double-$q$ (type 1) state is essentially the same state as the double-$q$ state observed in the triangular-lattice Heisenberg model in ref.~\cite{Okubo2}. The state can be  regarded as the superposition of the two spirals in the $xy$ component and  the lineally polarized spin density wave in the $z$ component. \textcolor{black}{Its real-space sublattice spin configurations corresponding to the spin structure factor shown in Figs.~11(a) and (b) are given in Figs.~\ref{double_type1_snap}(a) and (b) for the two sublattices.}

\subsection{\label{sec:The double-$q$ (type 2) state}D. The double-$q$ (type 2) state}
Next, we move to the second type of double-$q$ states, the double-$q$ (type 2) state. The corresponding intensity plots of the sublattice spin structure factors are given in  Figs.~\ref{double-q-matome}(c) and (d). Note that six peaks appear in the $xy$ component of the static spin structure factors shown in Figs. 11(c), \textcolor{black}{while two of them indicated in Fig.~11(c) are just broad peaks with weaker intensity than the other four peaks, about \textcolor{black}{52\%} weaker in intensity, spontaneously breaking the sixfold rotational symmetry. As can be seen from Figs.~\ref{double-q-matome}(c) and (d), the associated wavevectors run along the NN directions.} Hence, overall features of the static spin structure in this double-$q$ (type 2) state are similar to those in the double-$q$ (type 1) state. The real-space spin configuration, however, are very different. \textcolor{black}{We show in Figs.~\ref{double_type2_snap}(a) and (b) the real-space spin configurations in the double-$q$ (type 2) state, (a) for the sublattices I, and (b) for the sublattice II. In Fig.~\ref{double_type2_snap}(c)-(f),   spins at various sites on the sublattice I (c,d), and on the sublattice II (e,f) are reorganized with a common origin.} As can be seen from the figures, the double-$q$ (type 2) state has a coplanar structure in real space, spins lying on a plane containing the magnetic-field ($z$) axis, in contrast to the noncoplanar structure of the double-$q$ (type 1) state. 

\begin{figure}[t]
  \includegraphics[bb=230 80 552 492,width=4.5cm,angle=0]{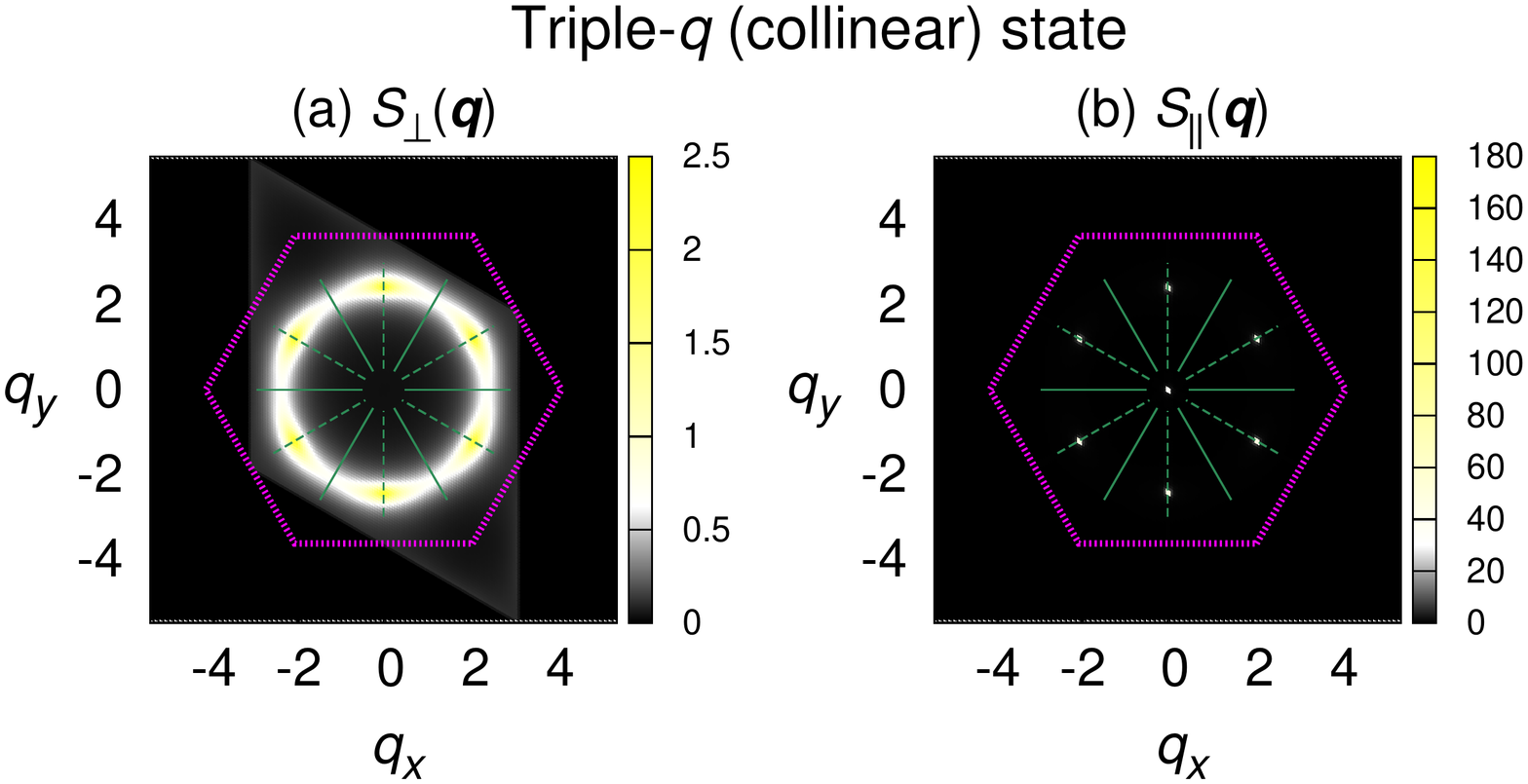}%-90->0, 3.5->4.5
 \caption{(Color online) The intensity plots of the sublattice spin structure factors in the wavevector ($q_x$, $q_y$) plane in the triple-$q$ (collinear) state; (a) the transverse component $S_{\perp}({\bf q})$, and (b) the longitudinal component $S_{\parallel}({\bf q})$. The parameters are $J_2/J_1=0.3$, $H/|J_1|=1.10$, $T/|J_1|=0.03617$ and $L=108$.
}
 \label{Sq_collinear}
\end{figure}

\begin{figure}[t]
  \includegraphics[bb=70 180 512 792, width=7.0cm,angle=0]{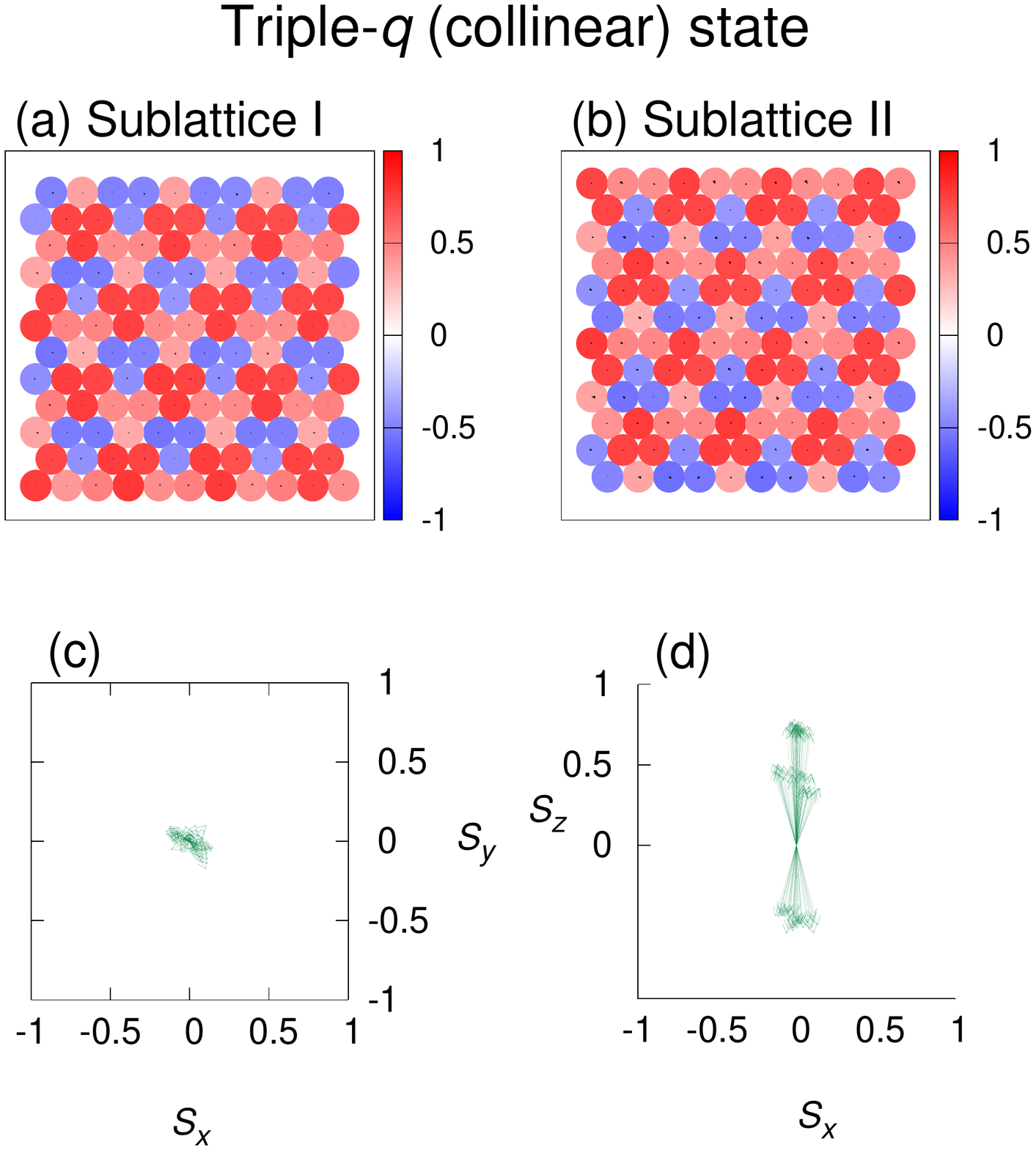}
 \caption{(Color online) Real-space sublattice spin configurations in the triple-$q$ (collinear) state obtained by the short-time average of 1000 MC steps, (a) for the sublattice I, and (b) for the sublattice II. The parameters are $J_2/J_1=0.3$, $H/|J_1|=1.5$, $T/|J_1|=0.03778$ and $L=200$. The $xy$ components of the spin are represented by the arrow, while the $z$ component is represented by the blue-to-red color scale. In (c) and (d), spins at various sites on the  sublattice I are reorganized with a common origin, (c) the top view in the ($S_x$, $S_y$) plane, and (d) the side view in the ($S_x$, $S_y$) plane. Essentially the same figure is obtained also for the sublattice II.
}
 \label{collinear_snap}
\end{figure}

\subsection{\label{sec:The double-$q$ (type 3) state}E. The double-$q$ (type 3) state}

 Still another type of double-$q$ state, the double-$q$ (type 3) state, is also possible, which actually occupies a rather wide region of the phase diagram of Fig.~\ref{phase_diagram}. The typical intensity plots of the double-$q$ (type 3) state are shown in Figs.~\ref{double-q-matome}(e) and (f). Unlike the double-$q$ states of type 1 or 2, the peaks of $S_{\parallel}({\bf q})$ appear at the same two-pair positions as those of $S_{\perp}({\bf q})$. Indeed, a closer inspection has revealed that the ordering wavevectors at the double-$q$ (type 3) state have features different from those of the other double-$q$ states (type 1 and type 2).  In Figs. 14(a) and (b), we show the intensity plots of both (a) $S_{\perp}({\bf q})$ and (b) $S_{\parallel}({\bf q})$, focused around one given spot located near the $q_y$ axis corresponding to the NN direction of the honeycomb-lattice, together with the curve of the degenerate ring. As can be seen from the figures, the peak positions of $S_{\perp}({\bf q})$ and of $S_{\parallel}({\bf q})$ differ somewhat, both being off lines from the NN direction in mutually opposite directions.  Hence, wavevectors  in the double-$q$ (type 3) state are actually composed of four individual wavevectors. Furthermore, both peak positions deviate even from the degenerate ring, the $|q^*|$-value being smaller than the degenerate-ring radius by about 2\% in $S_{\perp}({\bf q})$, and by about 4\% in $S_{\parallel}({\bf q})$. In fact, the deviation of the $|q^*|$-value from the degenerate ring is also seen for other states due to the effects of thermal fluctuations, though by the same amount between in $S_{\perp}({\bf q})$ and in $S_{\parallel}({\bf q})$.

 To further examine the nature of the ordering, 
we show in Figs.~\ref{double_type3_Sq_size_dep}(c) and (d) the size dependence of the $S_{\perp}({\bf q})$ and $S_{\parallel}({\bf q})$ peaks. As can be seen from the figure, the peak positions clearly differ in $|q|$ between in $S_{\perp}({\bf q})$ and in $S_{\parallel}({\bf q})$. On increasing the lattice size $L$, the peak of $S_{\perp}({\bf q})$ tends to sharpen, which seems consistent with the expected quasi-LRO of the $xy$ components: remember that the $xy$ spin components possess the U(1) (or SO(2)) symmetry around the $z$-axis under magnetic fields. A similar behavior is observed in the size dependence of the peak height of $S_{\parallel} ({\bf q})$ suggesting the existence of the quasi-LRO also in the longitudinal component.

 The real-space spin configurations in the double-$q$ (type 3) state are shown in Figs.~\ref{double_type3_snap}, (a-b) for the sublattice I, and (c-f) for the sublattice II. The \textcolor{black}{spin configurations} are shown in (a,c) and in (b,d), each set representing two different spatial regions in the same sample. As can be seen from the figures, the $xy$-spin components form interweaving vortex/antivortex lattice patterns. The state shown in Fig.~\ref{double_type3_snap}(a) looks like a periodic array of vortices as highlighted by the circled regions in the figure, while antivortex-like spin configurations are formed in the regions between the vortices. Likewise, the state shown in Fig.~\ref{double_type3_snap}(b) looks like a periodic array of antivortices, while vortex-like spin configurations are formed in the regions between the antivortices. The vortex-lattice-looking region on the sublattice I (Fig.~(a)) looks like the vortex-lattice also on the sublattice II with some phase shift (Fig.~(c)), and the antivortex-lattice-looking region on the sublattice I (Fig.~(b)) looks like the antivortex-lattice also on the sublattice II with some phase shift (Fig.~(d)). The reason why the vortex-lattice-looking region and the antivortex-lattice-looking region spatially alternates in the same sample is simply because the ${\bf q}$-value associated with the present vortex/antivortex lattice state is incommensurate with the underlying triangular sublattice, and the relative phase difference gradually modulates from lattice point to lattice point. Indeed, the $|{\bf q}|$-value associated with the present vortex/antivortex order is $|{\bf q}|\simeq 2.43$ as can be seen from Fig.~14(c), slightly off the threefold commensurate value of $|{\bf q}|= 4\pi/3\sqrt{3} \simeq 2.418 $.

 In Figs.~\ref{double_type3_snap}(e) and (f), spins on one sublattice at various sites are reorganized in the spin space with a common origin, a top view in the ($S_x$, $S_y$) plane in (e), and a side view in the ($S_x$, $S_z$) plane in (f). One can see from Figs.~\ref{double_type3_snap}(f) that the spin texture does not cover a whole sphere, only a half sphere being mapped like a half-skyrmion or ``meron''~\cite{Gross, Affleck, Moon}. The meron-like structure arises from the modulation of the spin $z$ component characterized by ${\bf q}^*$ observed in $S_{\parallel} ({\bf q})$. Hence, the sublattice spin structure in the double-$q$ (type 3) state is a \textcolor{black}{meron/antimeron-like} lattice rather than the vortex/antivortex lattice.

\subsection{\label{sec:The triple-$q$ (collinear) state}F. The triple-$q$ (collinear) state}

 We now move to the triple-$q$ state. We find only one type of triple-$q$ state in our model with $J_2/J_1$=0.3, the triple-$q$ (collinear) state. The corresponding sublattice spin structure factors are shown in Figs.~\ref{Sq_collinear}. $S_{\parallel}({\bf q})$ exhibits sharp peaks of equal intensities at all ${\bf q}^{*}$ wavevectors in the NN directions, keeping the $C_3$ lattice symmetry. The observation is consistent with the behavior of the $m_3$ order parameter shown in Fig.~\ref{temp-dep}(b). $S_{\perp}({\bf q})$ also exhibits peaks at the same positions as the ones of $S_{\parallel}({\bf q})$, while they remain very broad, suggesting only a weak SRO developed in the $xy$ spin component. The degenerate ring-like structure characteristic of the ring-liquid paramagnetic state is still clearly visible in $S_{\perp}({\bf q})$, suggesting the remanence of enhanced fluctuations similar to those of the ring-liquid state. As already mentioned, from the symmetry viewpoint, this collinear state is adiabatically identical with the $Z$ state observed in the triangular model~\cite{Okubo2}.

 A typical real-space spin configurations in the triple-$q$ (collinear) state are shown in Figs.~\ref{collinear_snap}, (a) for the sublattice I, and (b) for the sublattice II. As can be seen from the figures, \textcolor{black}{the spin $z$ component on each sublattice forms a super-triangular-lattice pattern, while the spin $xy$ components remain disordered, leading to the collinear spin ordering. The spin $z$-component configurations on the two sublattices turn out to be  essentially similar, their apparent difference borne by appropriate phase factors between the sublattices I and II. Namely, each of the triple-$q$ wavevectors, $q_1^*$, $q_2^*$ and $q_3^*$, possesses associated phase factors, $\alpha_{q_1^*}$, $\alpha_{q_2^*}$ and $\alpha_{q_3^*}$, which are not necessarily equal with each other, reflecting our choice of the unit cell indicated in Fig.~1, which apparently breaks the lattice $C_3$ symmetry. \textcolor{black}{Some more information of the MF level is given in the Supplemental Material.}}

 The triple-$q$ (collinear) state is realized in the phase diagram only at relative high temperatures and at intermediate fields, say, around $H/J_1\simeq 1.5$. In fact, the ordering behaviors at this intermediate field turns out to be quite rich, as can be seen from the $H$-$T$ phase diagram of Fig.~\ref{phase_diagram}. Namely, on decreasing the temperature from the ring-liquid paramagnetic state, the system first enters into the collinear triple-$q$ state, then into the noncollinear but coplanar double-$q$ state (double-$q$ (type 2) state), then into the noncoplanar double-$q$ state (double-$q$ (type 1) state), and eventually into the noncoplanar single-$q$ (NN) state at low enough temperatures.

\textcolor{black}{In concluding this section, we comment on our MF analysis very briefly. In order to clarify the origin of the various multiple-$q$ states observed in this section, we also perform a MF analysis following the methods of Reimers $et \ al$~\cite{Reimers} and of Okubo $et \ al$~\cite{Okubo, Okubo2}.  The details are given in the Supplement Material. Then, we succeed in obtaining all ordered state observed in our MC simulation, at least as the saddle-point solutions of the MF equations. It then turns out that the umbrella-type single-$q$ is the only stable solution and the all other states are just the saddle-point solutions, indicating that the umbrella-type single-$q$ state is always stabilized at the MF level. This observation highlights the crucial importance of fluctuations in stabilizing the various multiple-$q$ states in the present honeycomb-lattice system. The MF analysis also provides us some useful information about, {\it e.g.\/}, the running direction of the wavevector chosen from the degenerate ring, or provides us convenient compact expressions of the spin configurations describing the spin configurations in each multiple-$q$ state, {\it etc\/}, even when the multiple-$q$ states are not true stable states at the MF level.}

\section{\label{sec:other-values-of-J2}IV. Other values of $J_2/J_1$}

In the previous section, we have focused on the case of $J_2/J_1=0.3$. There, the ordering behavior has turned out to be quite rich, including a variety of multiple-$q$ states. In this section, we touch upon the ordering behavior of the model for other $J_2/J_1$-values in the range $1/6<J_2/J_1<0.5$, {\it i.e.\/}, $J_2/J_1=0.20, 0.25, 0.35$ and $0.45$. We find that, unlike the case of $J_2/J_1=0.30$, most of the $H$-$T$ phase diagrams consists of umbrella-type single-$q$ states, as shown in Figs.~\ref{other_J2_phase}, being qualitatively similar to the MF phase diagram \textcolor{black}{(see the Supplemental Material)}. Exceptions might be that, for $J_2/J_1=0.25$, the triple-$q$ (collinear) state appears in a narrow part of the phase diagram, and that, for $J_2/J_1=0.45$, two kinds of triple-$q$ states possessing different spin configurations from that in  the triple-$q$ (collinear) state, are stabilized. In fact, $J_2/J_1=0.45$ is close to the border line value $J_2/J_1=0.50$ at which the degenerate ring coincides with the first Brillouin zone (BZ) boundary, where underlying physics might be related to the BZ boundary. The two triple-$q$ states observed at $J_2/J_1=0.45$ are the ``3-1 triple-$q$ state'' and ``collinear 3-1 triple-$q$ state'' already reported in ref.~\cite{Rosales} for $J_2/J_1=0.5$. 
The 3-1 triple-$q$ state is not a collinear state different in nature from the triple-$q$ (collinear) state realized at $J_2/J_1=0.3$ and 0.25.
On the other hand, we find that the spin configuration in the collinear 3-1 triple-$q$ state, shown in \textcolor{black}{Appendix B}, can be described by \textcolor{black}{the same equation} as in the triple-$q$ (collinear) state \textcolor{black}{(see the eq. (33) in the Supplemental Material}). The apparent difference in their spin configurations comes only from the length of the ordering wavevectors on the associated degenerate ring.

 \textcolor{black}{We also comment that the emergence of an infinituple-$q$ ordered state called a ``ripple state'' was recently reported in ref.~\cite{TH_ripple} for the system of $J_2/J_1=0.18$  lying close to the AF phase boundary.}

\begin{figure}[t]
  \includegraphics[bb=100 0 612 792, width=12.0cm,angle=0]{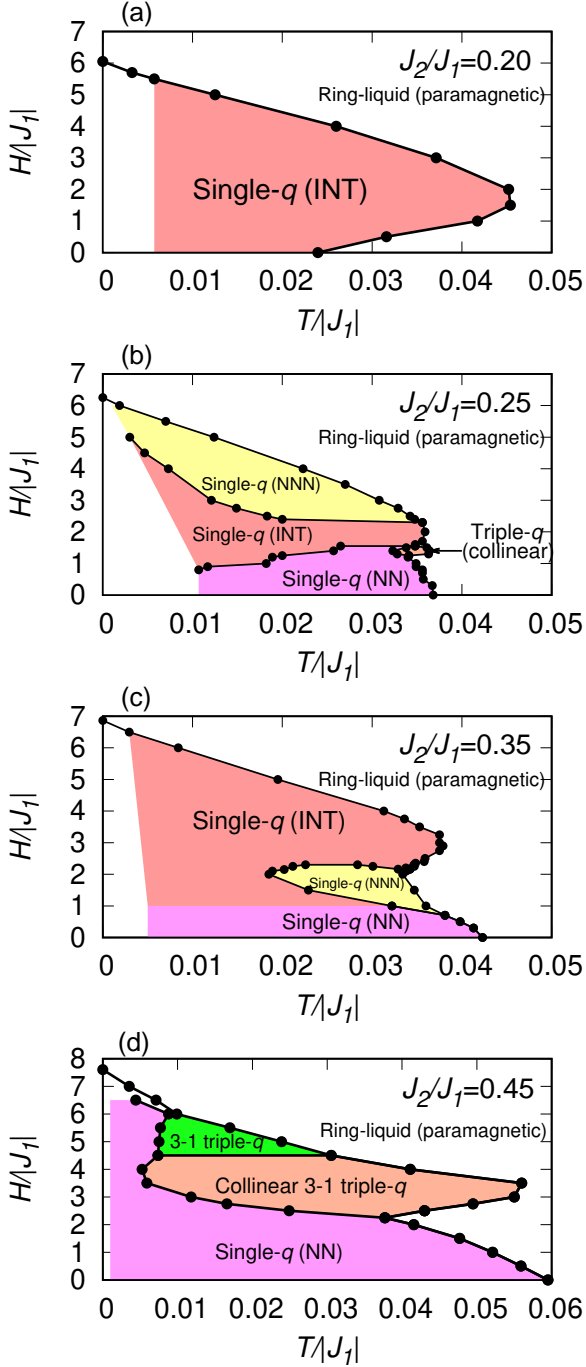}
 \caption{(Color online) The $H$-$T$ phase diagrams for $J_2/J_1=0.20, 0.25, 0.35$ and 0.45 obtained by MC simulations. Transition points denoted by black dots are determined from the specific-heat peak.
}
 \label{other_J2_phase}
\end{figure}

\section{\label{sec:Summary}V. Summary and Discussion}

In summary, we have investigated the ordering behaviors of the frustrated $J_1$-$J_2$ classical honeycomb-lattice Heisenberg AF under magnetic fields. Special attention has been paid to the case of $J_2/J_1=0.3$, which is located in the middle of the paramagnetic (ring-liquid) - helical phase boundary. The ring-like continuous degeneracy, and the resulting paramagnetic ring-liquid state provides a matrix of a rich variety of multiple-$q$ ordered states stabilized under fields. 

Via extensive MC simulations on the model, we have found a variety of multiple-$q$ states including the single-$q$, the double-$q$ and the triple-$q$ states, also including the noncoplanar, coplanar and collinear states. In contrast to the triangular-lattice case, the triple-$q$ skyrmion-lattice state is not stabilized. In fact, the obtained $H$-$T$ phase diagram turns out to differ considerably from that of the corresponding triangular-lattice model.
 
 For single-$q$ states, the umbrella-type and the fan-type single-$q$ states are found, the latter being stabilized only in high fields. The umbrella-type single-$q$ state occupies a wide region in the $H$-$T$ phase diagram, which might be further divided into several types of phases depending on the running direction of the associated $q$-vector. In the low-temperature limit, the $q$-vector running along the NN direction is preferred irrespective of the filed intensity for $J_2/J_1=0.3$ as is indicated by the low-temperature expansion calculation, whereas at higher temperatures the $q$-vector sometimes runs along the NNN direction or along the intermediate direction between NN and NNN. 

 In addition to these single-$q$ states, three distinct types of double-$q$ states, {\it i.e.\/}, the type 1, 2 and 3 double-$q$ states, are identified. The double-$q$ (type 1) state is similar to the double-$q$ state of the triangular-lattice model. It is a noncoplanar state with the double-$q$ structure in the $xy$ plane, forming the linearly polarized spin-density-wave along the $z$ direction. The double-$q$ (type 2) state is a coplanar state where spins lie on a plane containing the $z$-axis.  Particularly intriguing might be the double-$q$ (type 3) state, which corresponds to an interweaving meron/antimeron-like lattice state.

 The triple-$q$ state realized in the present honeycomb model at $J_2/J_1=0.3$ is the collinear triple-$q$ state in which the spin $z$-component forms a superlattice structure incommensurate with the underlying honeycomb lattice and the translational symmetry is spontaneously broken. This collinear state is adiabatically identical with the ``$Z$" state identified in the triangular-lattice model. In the latter state, the transverse spin correlation length turns out to be moderately long, which corresponds to the average domain size of the skyrmion (antiskyrmion) lattice domains. The triple-$q$ skyrmion-lattice state, which was observed to be stabilized in the triangular model in the vicinity of its $Z$ phase, turns out not to be stabilized in the present honeycomb model. 

\textcolor{black}{We also have investigated other $J_2/J_1$-values than $J_2/J_1$=0.3, to find that the dominant ordered state is an umbrella-type single-$q$ state except for the case near $J_2/J_1=0.5$. The richness of the $J_2/J_1=0.3$ phase diagram might be related to the fact that the $J_2/J_1$=0.3 point is located in the midst of the ring-liquid state in the $J_2/J_1$-$T$ phase diagram, as shown by Okumura et al in ref.~\cite{Okumura} (see its Fig.~9). In a wider parameter region of $0.2 \leq J_2/J_1 \leq0.45$, we have observed switching behaviors of the running directions of the critical wavevector in single-$q$ states as a function of the magnetic field and the temperature, which is likely to be a universal character of the honeycomb-lattice system.
}

 Finally, we wish to discuss possible implications of our present results to real magnets. One candidate material might be the $S=3/2$ honeycomb-lattice Heisenberg antiferromagnet ${\rm Bi_{3}Mn_{4}O_{12}(NO_{3})}$~\cite{Smirnova,Matsuda,SOkubo,Azuma, SOkubo2,Onishi}. Spin-liquid-like behavior was reported for this material in zero field, together with the field-induced antiferromagnetism. Further comprehensive experimental study of its in-field properties and the magnetic phase diagram might be interesting. This material actually consists of stacked honeycomb-bilayers, with the AF coupling between the two honeycomb layers. In comparing the present results with experiments especially under fields, care needs to be taken.

 The other candidate material might be a quantum bilayer kagome material ${\rm Ca_{10}Cr_{7}O_{28}}$, which was revealed to exhibit a spin-liquid-like behavior.~\cite{Balz1,Balz2} It was suggested that ${\rm Ca_{10}Cr_{7}O_{28}}$ might be modeled as a semi-classical honeycomb-lattice Heisenberg model with the ferromagnetic NN and the antiferromagnetic NNN interactions~\cite{Biswas,RHN}. This material exhibits a ring-liquid-like behavior with a characteristic ring-like pattern in the associated neutron scattering signal~\cite{Biswas,RHN}. Although $J_1$ is ferromagnetic in this material distinct from the one studied here, some of the ordering features may be common. Further study is desirable to clarify the situation.

\textcolor{black}{Finally, we wish to emphasize that the ring-like degeneracy and the resulting ring-liquid state could be a source of various exotic multiple-$q$ states. This is true not only in the present honeycomb-lattice system, but also in other systems with different lattice geometries, {\it e.g.\/}, a square-lattice system having a ring-like degeneracy in its ground-state was reported to exhibit a vortex crystal state in ref.~\cite{Seabra}. Another example might be a three-dimensional diamond-lattice system having a \textcolor{black}{surface-like} degenracy \cite{Bergman}, which was reported to give rise to different types of multiple-$q$ states\textcolor{black}{~\cite{Gao}}.}

 We hope that our present theoretical studies on a simple honeycomb model could provide a step toward the fuller understanding of rich ordering behaviors exhibited frustrated honeycomb magnets, or more generally, frustrated magnets possessing a massive ground-state degeneracy.

\subsection{\label{sec:ACKNOWLEDGMENTS}ACKNOWLEDGMENTS}
 One of the authors, T.S. is thankful to R. Pole, H. Yan and N. Shannon for fruitful discussions including the one about the recent new material ${\rm Ca_{10}Cr_{7}O_{28}}$. The authors are thankful to ISSP, the University of Tokyo and OIST for providing us with CPU time.  This study is supported by a Grant-in-Aid for Scientific Research No.~19K14665, No.~19K03740,  No.~25247064, No.~17H06137 and No.~15K17701.

\appendix
\renewcommand{\theequation}{A.\arabic{equation}}
\setcounter{equation}{0}

\section{\label{sec:Appendix}Appendix A. Low-Temperature Expansion}

In this appendix, we explain some of the details of the low-temperature expansion. Our low-temperature expansion is performed following the method described in refs.~\cite{Okumura, Bergman}. The partition function $Z$ of the model with the  Hamiltonian $H$
\begin{eqnarray}
Z=\int D {\bf S} e^{-\beta \mathcal{H}} \prod_{j=1}^N \delta [{\bf S}_j^2-1],
\end{eqnarray}
is evaluated by the low-temperature expansion from an arbitrary state on the degenerate ring as an unperturbed state. The fixed spin-length condition of the classical system requires that the ground state of the model is a single-$q$ state, and we assume as a ground state under magnetic fields an umbrella-like state given by
\begin{eqnarray}
\overline{\bf S}_{n}^a       &=& ( \sqrt{1-{m_z'}^2} {\rm cos} \theta_{n}^a, \sqrt{1-{m_z'}^2} {\rm sin} \theta_{n}^a, {m_z'}) \\
 \theta_{n}^a &=& {\bf q}^{*} \cdot {\bf r}_n + \alpha_{\bf q^*} \delta_{a{\rm II}},
\end{eqnarray}
where ${\bf r}_n$ is the position vector of the unit cell $n$, $a$ is the label for the two sublattices ($a={\rm I}$ or ${\rm II}$), ${\bf q^{*}}$ denotes the incommensurate spiral wavevector in the $xy$ plane, and $\alpha_{\bf q}$ denotes the sublattice phase difference as defined by \textcolor{black}{eq.~(6) of the Supplemental Material}. The spin longitudinal component $m'$ is obtained as

\begin{eqnarray}
{m'_z}=\frac{H}{\lambda_{{\bf q}^*}^+ - \lambda_{{\bf 0}}^-} ,
\end{eqnarray}
where the $\lambda_{{\bf q}}^{\pm}$ are the eigenvalues of the Hamiltonian given in \textcolor{black}{eq.~(10) of the Supplemental Material.}

Let us introduce the deviation vector $\bm{\pi}_n^{a}$, which satisfies $\bm{\pi}_n^{a} \perp \overline{\bf S}_{n}^a$. Then, we have 
\begin{eqnarray}
{\bf S}_{n}^{a} = \bm{\pi}_{n}^{a} + \overline{\bf S}_{n}^{a} \sqrt{1-\pi_j^2}.
\end{eqnarray}
The plane perpendicular to $\overline{\bf S}_{n}^{a}$ can be spanned by the two orthogonal unit vectors ${\bf e}_\perp$ and ${\bf e}_\perp \times \overline{\bf S}_{n}^{a}$, where
\begin{eqnarray}
{\bf e}_\perp = (-m_z'{\rm cos} \theta_{n}^a, -m_z'{\rm sin} \theta_{n}^a, \sqrt{1-{m_z'}^2}) .
\end{eqnarray}
We decompose the vector $\bm{\pi}_{n}^{a}$ as
\begin{eqnarray}
\bm{\pi}_{n}^{a} = \phi_n^{a} {\bf e}_\perp + \chi_n^{a} [{\bf e}_\perp \times \overline{\bf S}_{n}^{a}]
\end{eqnarray}
and expand the Hamiltonian up to the quadratic order both in $\chi$ and $\phi$. The partition function $Z$ can be written as
\begin{eqnarray}
Z=\int \prod_{n,a} d \phi_n^{a} d \chi_n^{a} e^{-\beta \mathcal{H}}.
\end{eqnarray}
which can be evaluated by the Gaussian integrals. Neglecting the terms independent of the critical wavevector ${\bf q}^{*}$, we finally get the following expression of the ${\bf q}^{*}$-dependent part of the free energy density,

\begin{eqnarray}
F({\bf q}^{*})/T &\sim&  \int d{\bf q}  \{ {\rm ln}[-(W_{\rm I,I}+|W_{\rm I,II}|) ] \nonumber \\
                  &+&                   {\rm ln}[-(W_{\rm I,I}-|W_{\rm I,II}|) ] \},
\label{lowT_F}
\end{eqnarray}
where
\begin{eqnarray}
&&W_{\rm I,I}({\bf q}^{*},{\bf q})= 2J_2 \{ \nonumber \\
                          & &[(1-{m_z'}^2){\rm cos}({\bf q}^{*}\cdot {\bf a}_x)+{m_z'}^2]{\rm cos}({\bf q} \cdot {\bf a}_x) \nonumber \\
                          &+& [(1-{m_z'}^2){\rm cos}({\bf q}^{*}\cdot {\bf a}_y)+{m_z'}^2]{\rm cos}({\bf q} \cdot {\bf a}_y) \nonumber \\
                          &+& [(1-{m_z'}^2){\rm cos}({\bf q}^{*}\cdot ({\bf a}_x-{\bf a}_y))+{m_z'}^2]{\rm cos}({\bf q} \cdot ({\bf a}_x-{\bf a}_y)) \} \nonumber \\
                                     &-& \lambda_{+}({\bf q}^{*}),
\label{lowT_I_I}
\end{eqnarray}

\begin{eqnarray}
W_{\rm I,II}({\bf q}^{*}, {\bf q}) &=& J_1 \{ \nonumber \\
       & &        [(1-{m_z'}^2) {\rm cos}                                \alpha_{{\bf q}^{*}}   + {m_z'}^2] \nonumber \\
       &+&        [(1-{m_z'}^2) {\rm cos} ({\bf q}^{*} \cdot {\bf a}_{x} -\alpha_{{\bf q}^{*}} ) + {m_z'}^2 ] e^{i {\bf q}\cdot {\bf a}_x} \nonumber \\
       &+&        [(1-{m_z'}^2) {\rm cos} ({\bf q}^{*} \cdot {\bf a}_{y} -\alpha_{{\bf q}^{*}} ) + {m_z'}^2 ] e^{i {\bf q}\cdot {\bf a}_y} \}. \nonumber \\
       & &
\label{lowT_I_II}
\end{eqnarray}
When $m_z^{\prime}$=0, these equations reduce to the ones given in the Appendix of ref.~\cite{Okumura}.

\section{\label{sec:Appendix B}Appendix B. The collinear 3-1 triple-$q$ state}
\begin{figure}[h]
  \includegraphics[bb=90 0 612 590, width=6.6cm,angle=0]{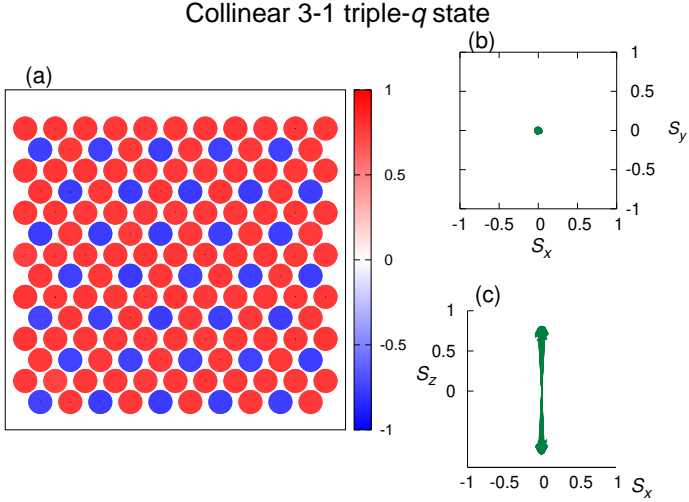}%-90->0
 \caption{(Color online) Real-space sublattice spin configuration in the collinear 3-1 triple-$q$ state obtained by the short-time average of 1000 MC steps. The parameters are $J_2/J_1=0.45$, $H/|J_1|=3.0$, $T/|J_1|=0.04006$ and $L=120$ under periodic boundary conditions. (a) The $xy$ components of the spin are represented by the arrow, while the $z$ component is represented by the blue-to-red color scale. In (b) and (c), spins at various sites on a given sublattice are reorganized with a common origin, (b) the top view in the ($S_x$, $S_y$) plane, and (c) the side view in the ($S_x$, $S_z$) plane.
}
 \label{collinear_type2_snap}
\end{figure}

We present the real-space sublattice spin configuration for the collinear 3-1 triple-$q$ state observed for $J_2/J_1=0.45$, as given in Fig.~\ref{collinear_type2_snap}. We find that this spin configuration can be reproduced by \textcolor{black}{the eq.~(35) of the Supplemental Material, that means, the same equation of the triple-$q$ (collinear) state observed for the $J_2/J_1=0.3$}, with the ordering wavevectors on the degenerate ring corresponding to $J_2/J_1=0.45$.

\newpage
\clearpage
 %%%%%%%%%%%%%%%%%%%%%%%%%%%%%%%%%%%%%%%%%%%%
 %Supplement Material
%%%%%%%%%%%%%%%%%%%%%%%%%%%%%%%%%%%%%%%%%%%%%

%\usepackage{here}
        \renewcommand{\thetable}{S\arabic{table}}%
        \setcounter{figure}{0}
        \renewcommand{\thefigure}{S\arabic{figure}}%
\renewcommand{\thesection}{\Roman{section}}
\makeatother
\setcounter{figure}{0} 
\setcounter{equation}{0}
\renewcommand{\theequation}{\arabic{equation}}

\onecolumngrid
\begin{center} {\bf \large Multiple-$q$ states of the $J_1$-$J_2$ classical honeycomb-lattice Heisenberg antiferromagnet under magnetic fields -Supplemental Material-} \end{center}
\vspace{0.5cm}
\twocolumngrid

\author{Tokuro Shimokawa} 
\email{tokuro.shimokawa@oist.jp}
\affiliation{Okinawa Institute of Science and Technology Graduate University, Onna, Okinawa, 904-0495, Japan}
\author{Tsuyoshi Okubo}
\affiliation{Department of Physics, University of Tokyo, Tokyo 113-0033, Japan }
\author{Hikaru Kawamura}
\affiliation{Department of Earth and Space Science, Graduate School of Science, Osaka University, Toyonaka, Osaka 560-0043, Japan}

\date{\today}
\maketitle

\section{\label{sec:Mean-field analysis}Mean-field analysis}

In order to get further insights into the possible origin of the various multiple-$q$ states observed in our MC simulations, we also perform a mean-field analysis following the methods of Reimers $et \ al$~\cite{Reimers_s} and of Okubo $et \ al$~\cite{Okubo_s, Okubo2_s}. \textcolor{black}{The MF analysis has an obvious limitation of neglecting fluctuations. Yet, it still gives us useful information about the stability mechanism of various multiple-$q$ states observed in our MC simulations. Furthermore, the comparison of the MF results with the MC results could reveal the importance of fluctuations in the stabilizing mechanism of each phase. In fact, all multiple-$q$ state observed in our MC simulations are also obtained in our MF analysis, at least as \textcolor{black}{the saddle-point solutions.} Comparison of the MF free energy of each state and the dependences on the temperature and the magnetic field help us understand the stability of each phase.}

 The Landau free energy per unit cell of the honeycomb-lattice Heisenberg model (1) is given up to quartic order by
\begin{eqnarray}
2F/N &=& -2T {\rm ln}4 \pi -\sum_{a} {\bf H} \cdot {\bf B}_{{\bf q}={\bf 0}}^{a} \nonumber \\
     &+& \frac{1}{2} \sum_{{\bf q}}\sum_{ab}  {\bf B}_{{\bf q}}^{a} \cdot  {\bf B}_{-{\bf q}}^{b} (3T \delta_{ab}- J_{{\bf q}}^{ab}) \nonumber \\
    &+& \frac{9T}{20} \sum_{a} {\sum_{\{{\bf q}\}}}' ( {\bf B}_{{\bf q}_1}^{a} \cdot  {\bf B}_{{\bf q}_2}^{a})( {\bf B}_{{\bf q}_3}^{a} \cdot  {\bf B}_{{\bf q}_4}^{a}),
\label{free1}
\end{eqnarray}
where $\delta_{ab}$ is a Kronecker delta, and ${\bf B}_{\bf q}^{a}$ is the order parameter corresponding to the Fourier magnetization of the sublattice $a$ ($a={\rm I,II}$) defined by
\begin{eqnarray}
{\bf B}_{\bf q}^{a} &=& \langle {\bf S}_{\bf q}^{a} \rangle, \\
{\bf S}_{\bf q}^{a} &=& \frac{2}{N} \sum_{n} {\bf S}_{n}^{a} {\rm exp} (- i{\bf q}\cdot {\bf r}_{n}),
\end{eqnarray}
where ${\bf S}_{n}^{a}$ is the spin at the unit cell $n$ belonging to the sublattice $a$, and ${\bf r}_{n}$ is the position vector of the unit cell $n$. \textcolor{black}{We take the unit cell as indicated by the green box in Fig.~1 of the main text, and take ${\bf r}_{n}$ to be the position vector of the site on the sublattice I.} The sum ${\sum_{\{{\bf q}\}}}'$ is taken under the constraint ${\bf q}_1 + {\bf q}_2 + {\bf q}_3+ {\bf q}_4=\textcolor{black}{{\bf 0}}$ up to reciprocal lattice vectors.  The $J_{{\bf q}}^{ab}$ is the Fourier transform of the exchange interaction between the sublattices $a$ and $b$, and the matrix $J_{\bf q}$ is given explicitly by 

\begin{eqnarray}
%\[
  J_{\bf q} = \left(
    \begin{array}{ccc}
      J_{\bf q}^{\rm I,I} & J_{\bf q}^{\rm I,II}  \\
      J_{\bf q}^{\rm II,I} & J_{\bf q}^{\rm II,II}  
    \end{array}
  \right)
%\]
\label{Jq_matrix}
\end{eqnarray}
with 

\begin{eqnarray}
J_{\bf q}^{\rm I,I} &=& J_{\bf q}^{\rm II,II} = 2 J_2 B, \nonumber \\
J_{\bf q}^{\rm I,II} &=& (J_{\bf q}^{\rm II,I})^*=|J_1| A {\rm exp}(\mathrm{i}\alpha_{\bf q}), \nonumber \\
A &=& (3+2B)^{1/2}, \\
B &=& {\rm cos}(\tilde{q}_x) + {\rm cos}(\tilde{q}_y) + {\rm cos}(\tilde{q}_x-\tilde{q}_y), \nonumber
\end{eqnarray}
where $\alpha_{\bf q}$ is defined by

\begin{eqnarray}
{\rm cos}(\alpha_{\bf q}) &=& {\rm sgn} (J_1) (1+{\rm cos}(\tilde{q}_x) +{\rm cos}(\tilde{q}_y))/A , \nonumber \\
{\rm sin}(\alpha_{\bf q}) &=& {\rm sgn} (J_1) ({\rm sin}(\tilde{q}_x)+ {\rm sin}(\tilde{q}_y))/A , \nonumber \\
\label{eqalpha}
\end{eqnarray}
with $\tilde{q}_{x(y)} \equiv {\bf q}\cdot {\bf a}_{x(y)}$. 

 One can diagonalize the quadratic term in eq.~(\ref{free1}) via a unitary matrix $U_{\bf q}$, with the $i$-th eigenvalue $\lambda_{\bf q}^{i}$,

\begin{eqnarray}
\sum_{b} J_{\bf q}^{ab} U_{\bf q}^{bi}=\lambda_{\bf q}^{i} U_{\bf q}^{ai}.
\end{eqnarray}
Transforming the order parameter to normal modes ${\bf \Phi}_{\bf q}^{i}$,

\begin{eqnarray}
{\bf B}_{\bf q}^{a}=\sum_{i} U_{\bf q}^{ai} {\bf \Phi}_{\bf q}^{i},
\end{eqnarray}
the Landau free energy $F$ of eq.~(\ref{free1}) can be rewritten as

\begin{eqnarray}
2F/N &=& -2T {\rm ln}4 \pi -\sum_{a,i} U^{ai}_{{\bf q}={\bf 0}} {\bf H} \cdot {\bf \Phi}_{{\bf q}={\bf 0}}^{i} \nonumber \\
    &+& \frac{1}{2} \sum_{{\bf q},i}  {\bf \Phi}_{{\bf q}}^{i} \cdot  {\bf \Phi}_{-{\bf q}}^{i} (3T- \lambda_{{\bf q}}^{i}) \nonumber \\
    &+& \frac{9T}{20} \sum_{ijkl} {\sum_{\{{\bf q}\}}}' ( {\bf \Phi}_{{\bf q}_1}^{i} \cdot  {\bf \Phi}_{{\bf q}_2}^{j})( {\bf \Phi}_{{\bf q}_3}^{k} \cdot  {\bf \Phi}_{{\bf q}_4}^{l}) \nonumber \\
    &\times& \sum_{a} U_{{\bf q}_1}^{ai}  U_{{\bf q}_2}^{aj}  U_{{\bf q}_3}^{ak}  U_{{\bf q}_4}^{al},
\label{free2}
\end{eqnarray}
where $i,j,k$ and $l$ are eigen-mode indices.

The matrix $J_{\bf q}$ of eq.~(\ref{Jq_matrix}) can easily be diagonalized, yielding the two eigenvalues

\begin{eqnarray}
\lambda_{\bf q}^{\pm} = J_{\bf q}^{\rm I,I} \pm |J_{\bf q}^{\rm I,II}| = 2 J_2 B \pm |J_1| \sqrt{3+ 2B},
\label{ring}
\end{eqnarray}
with the corresponding eigenvectors given by

\begin{eqnarray}
{\bf U}_{\bf q}^{\pm}= 
   \left(
    \begin{array}{cc}
      U^{\rm I\pm}_{\bf q}   \\
      U^{\rm II\pm}_{\bf q}  
    \end{array}
  \right)
=
\frac{1}{\sqrt{2}} 
   \left(
    \begin{array}{cc}
      1   \\
      \pm {\rm exp}(- \mathrm{i}\alpha_{\bf q})  
    \end{array}
  \right),
\label{eigenvec}
\end{eqnarray}
where $\alpha_{\bf q}$ actually represents an angle representing the phase difference between the two sublattices, satisfying the relation \textcolor{black}{ $\alpha_{-{\bf q}}=-\alpha_{\bf q}$}. The $\alpha_{\bf q}$ term arises from the non-Bravais nature of the honeycomb lattice, which does not exist in the triangular-lattice case \cite{Okubo2_s}. Indeed, $\alpha_{\bf q}$ plays an important role in understanding the nature of multiple-$q$ states in our honeycomb-lattice model as will be discussed below.

\begin{figure}[h]
  \includegraphics[bb=50 50 612 760, width=9.0cm,angle=0]{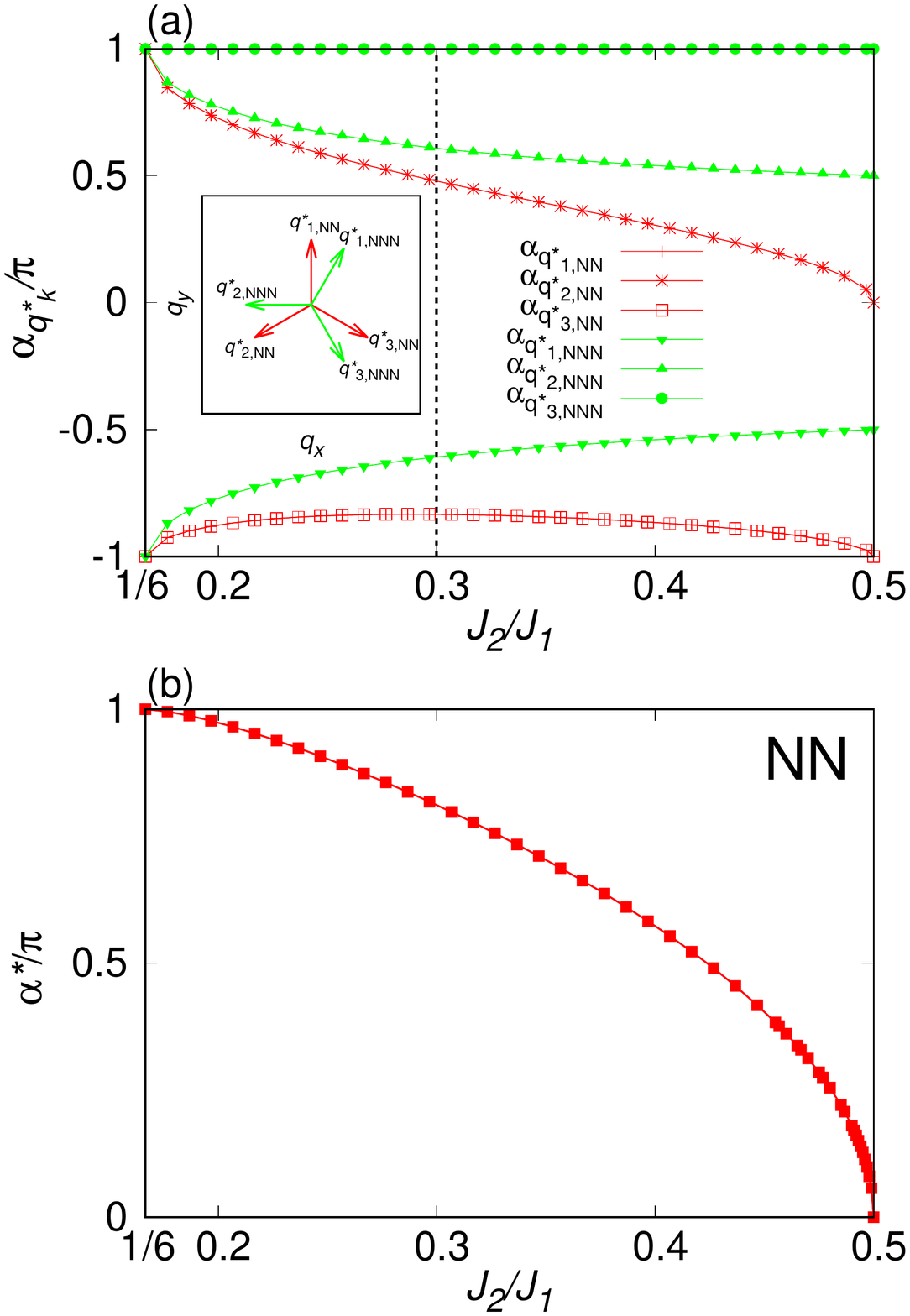}
 \caption{(Color online) \textcolor{black}{(a) The $J_2/J_1$-dependence of the sublattice phase differences $\alpha_{{\bf q}^*}$s for the critical wavevectors on the degenerate ring.  \textcolor{black}{As representative directions, we choose the wavevectors running along the NN (red) or the NNN (green) directions. Typical three vectors in the NN (NNN) directions are displayed in the inset of (a) as red (green) arrows.}  (b) The sum of the sublattice phase differences, $\alpha^*=\alpha_{{\bf q}^{*}_{1,{\rm NN}}}+ \alpha_{{\bf q}^{*}_{2,{\rm NN}}}+\alpha_{{\bf q}^{*}_{3,{\rm NN}}}$+$2\pi$, in the case where the $q$-vectors run along the NN direction.}}
 \label{sum_of_alpha_NN2}
\end{figure}

 As we could see \textcolor{black}{in the section III of the main text} for $J_2/J_1=0.3$, most of multiple-$q$ states observed in MC calculations have critical wavevectors running along the NN or NNN directions. According to eq.~(\ref{eqalpha}), \textcolor{black}{we can get the information about the $\alpha$.}
\textcolor{black}{In Fig.~\ref{sum_of_alpha_NN2}(a), we show the $J_2/J_1$-dependence of the $\alpha_{{\bf q}^*}$ values for the critical wavevector on the degenerate ring, especially the ones running in the NN and NNN directions.  Note that, reflecting our choice of the unit cell, $\alpha_{q_1^*}$,  $\alpha_{q_2^*}$ and $\alpha_{q_3^*}$ are no longer equivalent.
 In Fig.~\ref{sum_of_alpha_NN2}(b), the sum of the phase differences in the case of the NN directions, $\alpha^*=\alpha_{{\bf q}_{1}^*} +\alpha_{{\bf q}_{2}^*} + \alpha_{{\bf q}_{3}^*}$, is shown.}
\textcolor{black}{From these figures, the $\alpha$-values for the $J_2/J_1$=0.30} on the degenerate ring are $\alpha_q \sim \mp 0.834\pi$ or $\sim \pm 0.479\pi$ for the NN directions; the former value is for the wavevector along (0, $\pm$1) and ($ \pm \frac{\sqrt{3}}{2}$, $ \mp \frac{1}{2}$) directions ($\pm q_{1,{\rm NN}}^*$ and $\pm q_{3,{\rm NN}}^*$ directions in Fig.~\ref{sum_of_alpha_NN2}), while the latter one is for the wavevector along ($\mp \frac{\sqrt{3}}{2}$, $ \mp \frac{1}{2}$) direction ($\pm q_{2,{\rm NN}}^*$ direction in Fig.~\ref{sum_of_alpha_NN2}). For the NNN directions, the corresponding $\alpha$-value are $\sim \mp 0.608 \pi$ or $\pm \pi$; the former value is for the wavevector along ($\pm \frac{1}{2}$, $\pm \frac{\sqrt{3}}{2}$) and ($\pm$1, 0) directions ($\pm q_{1,{\rm NNN}}^*$ and ($\pm q_{3,{\rm NNN}}^*$ directions in Fig.~\ref{sum_of_alpha_NN2}), while the latter one is for the wavevector along ($\pm \frac{1}{2}$, $\mp \frac{\sqrt{3}}{2}$) direction ($\pm q_{2,{\rm NNN}}^*$ direction in Fig.~\ref{sum_of_alpha_NN2}). 

 One sees from eq.~(\ref{free2}) that, around the transition temperature $T_c$, the leading contribution comes from the maximum eigenvalue $\lambda_{{\bf q}^{*}}^{+} \equiv{\rm max} _{{\bf q}} \{ \lambda_{\bf q}^{+}\}$, where the maximum eigenvalue always lies in the $\lambda_{\bf q}^{+}$ branch. The transition temperature is then given by $T_c=\frac{1}{3} \lambda_{{\bf q}^{*}}^{+}$. 

 For the case of $1/6<J_1/J_2<0.5$ of our interest, the corresponding wavevector ${\bf q}^{*}$ is on the degenerate ring around the \textcolor{black}{origin in the sublattice wavevector space}, given by

\begin{eqnarray}
{\rm cos}(\tilde{q}_x^{*}) + {\rm cos}(\tilde{q}_y^{*}) + {\rm cos}(\tilde{q}_x^{*}-\tilde{q}_y^{*}) = \frac{J_1^2}{8J_2^2}-\frac{3}{2}.
\label{ring2}
\end{eqnarray}
The curves in \textcolor{black}{Fig.~2 of the main text} representing the degenerate ring was drawn by using this equation.

In fact, because of an infinite degeneracy associated with this ring, the relevant $q^*$'s are still largely undetermined at the quadratic level. Since we consider only $\lambda_{\bf q}^{+}$ branch, the order parameter $m^2$ is given by

\begin{eqnarray}
 m^{2} = \sum_{{\bf q}^{*}} |{\bf \Phi}_{{\bf q}^{*}}^{+}|^2.
\label{opar}
\end{eqnarray}

 In the Landau expansion, we neglect the contribution from $\lambda_{\bf q}^{-}$ branch, except for the uniform one, {\it i.e.\/}, the ${\bf q}={\bf 0}$ mode which belongs to the $\lambda_{{\bf q}}^{-}$ branch. It plays an obviously important role under magnetic fields since it describes the uniform magnetization, and we treat the uniform ${\bf q}={\bf 0}$ mode from the $\lambda_{\bf q}^{-}$ branch in considering the free energy (\ref{free2}).

 Then, the Landau-type free energy $F$ is reduced to

\begin{eqnarray}
2F/N &=& -2T {\rm ln}4 \pi -\sqrt{2}  H  \Phi_{\bf 0}^{-} + \frac{1}{2} \sum_{{\bf q}^{*}}  |{\bf \Phi}_{{\bf q}^{*}}^{+}|^2 (3T- \lambda_{{\bf q}^{*}}^{+}) \nonumber \\
    &+& \frac{1}{2}|{\bf \Phi}_{{\bf 0}}^{-}|^2 (3T- \lambda_{\bf 0}^{-}) + f_4,
\label{free3}
\end{eqnarray}
where the $f_4$ is the quartic term given by

\begin{eqnarray}
f_4 &=&\frac{9T}{20} {\sum_{\{i,{\bf q}\}}}'' ( {\bf \Phi}_{{\bf q}^{}_1}^{i_1} \cdot  {\bf \Phi}_{{\bf q}^{}_2}^{i_2})( {\bf \Phi}_{{\bf q}^{}_3}^{i_3} \cdot  {\bf \Phi}_{{\bf q}^{}_4}^{i_4}) \nonumber \\
    &\times& \sum_{a={\rm I,II}} U_{{\bf q}^{}_1}^{ai_1}  U_{{\bf q}^{}_2}^{ai_2}  U_{{\bf q}^{}_3}^{ai_3}  U_{{\bf q}^{}_4}^{ai_4}.
\label{f4}
\end{eqnarray}
where the summation over $\{i, {\bf q}\}$ is taken over all possible combinations of ($+,{\bf q}_1^{*}$), ($+,{\bf q}_2^{*}$), ($+,{\bf q}_3^{*}$) and ($-,{\bf 0}$) allowing multiplicity, with the wavevector conservation. Under the ring-like infinite degeneracy, there are infinitely many ways at the MF level in choosing the combination of ${\bf q}^{*}$ vectors at the quartic term (\ref{f4}). In fact, fluctuations often lift this degeneracy via the mechanism of order by disorder. 

 Under these assumptions, the real-space spin configurations at the MF level are given by
\textcolor{black}{
\begin{eqnarray}
{\bf S}_{n}^{a} = \sum_{k} \{ {\rm Re[} {\bf \Phi}_{{\bf q}^{*}_{k}}^{+} {\rm ] cos}(\Delta_{n,k})+{\rm Im[} {\bf \Phi}_{{\bf q}^{*}_{k}}^{+} {\rm ] sin}(\Delta_{n,k})  \} + {\bf \Phi}_{\bf 0}^{-}, \nonumber \\
\end{eqnarray}
with
\begin{eqnarray}
\Delta_{n,k}={\bf q}^{*}_{k} \cdot {\bf r}_{n} + \theta_k +\alpha_{{\bf q}^{*}_{k}} \delta_{a {\rm II}}
\end{eqnarray}
where ${\rm Re[} {\bf \Phi}_{{\bf q}^{*}_{k}}^{+} {\rm ]}$ and ${\rm Im[} {\bf \Phi}_{{\bf q}^{*}_{k}}^{+} {\rm ]}$ are the real and imaginary parts of ${\bf \Phi}_{{\bf q}^{*}_{k}}^{+}$, respectively, and  $\theta_k$ is an arbitrary phase factor of each normal mode ${\bf \Phi}_{{\bf q}^{*}_{k}}^{+}$.}

 The umbrella-type single-$q$ state associated with an eigenvector ${\bf U}_{\bf q}^{+}$ satisfies the fixed spin-length condition $|{\bf S}_i^{a={\rm I}}|=|{\bf S}_i^{a={\rm II}}|=1$, and can thereby be realized even at $T=0$. By contrast, the emergence of multiple-$q$ states in the ground state is difficult, since the state cannot satisfy the fixed-spin-length condition~\cite{Leonov_s}. Even so, the multiple-$q$ states could be stabilized at finite temperatures due to thermal fluctuations as observed in Sec. III \textcolor{black}{of the main text} and in some other works~\cite{Okubo2_s, Rosales_s, Seabra_s}.

 In the following subsections, with the hope to describe various ordered states observed in our MC simulations, we search for typical solutions giving extreme values of the MF equation. \textcolor{black}{Note that, in finding \textcolor{black}{the saddle-point solutions} of our MF free energy, we use the knowledge obtained from the static spin structure factors computed by our MC simulations, {\it e.g.\/}, the number of critical wavevectors in the $xy$ and $z$ components \textcolor{black}{and their relative relations.}}

\subsection{\label{sec:The single-$q$ states}A. The single-$q$ states}
We first consider the single-$q$ states composed of independent critical modes ${\bf \Phi}_{\pm {\bf q}^{*}}^{+}$ and a uniform mode ${\bf \Phi}_{\bf 0}^{-}$.  For the single-$q$ state, there are two typical solutions, one with the umbrella-type spin configuration and the other with the fan-type configuration.

The real-space sublattice spin configuration of the umbrella-type single-$q$ state is given by
\begin{eqnarray}
\langle {\bf S}_{n}^{a} \rangle &\propto&  
   \left(
    \begin{array}{cc}
     \frac{m}{2} {\rm cos}\Delta_n \\
     \frac{m}{2} {\rm sin}\Delta_n \\
      \Phi_{\bf 0}^{-}
    \end{array}
  \right),
\label{config_umbrella}
\end{eqnarray}
with
\begin{eqnarray}
\Delta_n \equiv {\bf q}^{*}\cdot {\bf r}_n + \theta +\alpha_{{\bf q}^{*}} \delta_{a{\rm II}} ,
\label{config_umbrella2}
\end{eqnarray}
where $\theta$ is an arbitrary phase factor, $\alpha_{{\bf q}^{*}}$ a phase difference between the sublattices I and II determined by eq.~(\ref{eqalpha}), and $m$ the amplitude defined by $m^2=|{{\Phi}_{ {\bf q}^{*}}^{+}}|^2 + |{{\Phi}_{ -{\bf q}^{*}}^{+}}|^2$. 

 In the umbrella-type structure, the $xy$-components form a spiral while the $z$-component forms a uniform ${\bf q}={\bf 0}$ magnetization along the applied magnetic field $H$. We find that the single-$q$ MF solution is consistent with all types of the umbrella-type single-$q$ states including the NN, NNN, and INT states as observed in the MC simulation.

The corresponding $f_4$ is given by
\begin{eqnarray}
f^{\rm umbrella}_{4} &=&
                  \frac{9T}{20} \bigl[ \frac{m^4}{2} + \frac{1}{2} {{\Phi}_{\bf 0}^{-}}^4 + m^2 {{\Phi}_{\bf 0}^{-}}^2 \bigr].%,\\
\label{free_single}
\end{eqnarray}
The parameters $m$ and ${\Phi}_{\bf 0}^{-}$ are the variational parameters to be determined to minimize the total free energy $F$ for a given set of temperature $T$, magnetic field $H$ and $J_2/J_1$ values. At the MF level, the free energy of the umbrella-type single-$q$ state does not depend on the direction of the critical wavevector ${\bf q}^{*}$ as long as it is on the degenerate ring even under magnetic fields. In this sense, in order to determine the selected ${\bf q}^{*}$ direction, we need to go beyond the MF level. In fact, thermal fluctuations select the NN direction, which we \textcolor{black}{discussed} in Sec.~III.~B by employing the low-temperature expansion.

 In the fan-type single-$q$ state, in contrast to the umbrella-type single-$q$ state, the state is coplanar where the spins lie on a common plane including the magnetic-field axis, say, $xz$-plane, whose spin configuration is given by
\begin{eqnarray}
\langle {\bf S}_{n}^{a} \rangle &\propto&  
   \left(
    \begin{array}{cc}
      \frac{m}{2} ({\rm cos}(\Delta_n)
     + {\rm sin}(\Delta_n))    \\
     0  \\
      \Phi_{\bf 0}^{-}
    \end{array}
  \right) ,
\label{config_fan}
\end{eqnarray}
and all configurations obtained by rotating it around the $z$-axis: recall a $U(1)$ spin-rotation symmetry around the $z$-axis. The $f_4$ of this fan-type single-$q$ state is given by
\begin{eqnarray}
f^{\rm fan}_{4} =\frac{9T}{20} \bigl[ \frac{3m^4}{4} + \frac{1}{2} {{\Phi}_{\bf 0}^{-}}^4 + m^2 {{\Phi}_{\bf 0}^{-}}^2 \bigr].
\label{free_single2}
\end{eqnarray}
The free energy of the fan-type single-$q$ state turns out to be always higher than that of the umbrella-type single-$q$ state. Hence, at the MF level, the fan-type single-$q$ state has no chance to be stabilized as a stable state. As revealed by our MC simulations, however, thermal fluctuations stabilize the fan-type single-$q$ state at finite temperatures in the high-field region. \textcolor{black}{Again, the free energy does not depend on the direction of the wavevector, without any constraint imposed on the running direction of the $q^*$-vector. This MF observation is consistent with our MC observation that the intermediate direction, instead of the high-symmetric NN and NNN directions, is realized in the fan-type single-$q$ state.}

 %Again, the free energy does not depend on the direction of the wavevector, with no constraint imposed on the running direction of the $q^*$-vector at the MF level. Such a MF observation is fully consistent with the MC observation that the $q^*$-vector runs along the INT direction in the fan-type single-$q$ state.

\subsection{\label{sec:The double-$q$ (type 1) and (type 2) states}B. The double-$q$ (type 1) and (type 2) states}
We next consider the double-$q$ states composed of two independent critical modes ${\bf \Phi}_{\pm{\bf q}^{*}_{1}}^{+}$, ${\bf \Phi}_{\pm{\bf q}^{*}_{2}}^{+}$ in the $xy$ plane and another critical mode ${\bf \Phi}_{\pm{\bf q}^{*}_{3}}^{+}$ along the $z$ axis with an additional uniform mode ${\bf \Phi}_{\bf 0}^{-}$.
 For this kind of double-$q$ state, we find two typical solutions, each corresponding to the double-$q$ (type 1) state and the double-$q$ (type 2) state observed in our MC simulations.

The real-space sublattice spin configuration corresponding to the double-$q$ (type 1) state is given by
\begin{eqnarray}
\langle {\bf S}_{n}^{a} \rangle &\propto& 
   \left(
    \begin{array}{cc}
     \frac{m_{xy}}{2} \bigl[ {\rm cos} (\Delta_{n,1}) +  {\rm cos} (\Delta_{n,2})  \bigr] \\
     \frac{m_{xy}}{2} \bigl[ {\rm sin} (\Delta_{n,1}) -  {\rm sin} (\Delta_{n,2})  \bigr] \\
           m_{z}             {\rm cos}(\Delta_{n,3})  +  \Phi_{\bf 0}^{-}
    \end{array}
  \right) ,
\end{eqnarray}
%\nonumber \\
%&{\rm with}& \ 
with
\begin{eqnarray}
\Delta_{n,k} \equiv {\bf q}^{*}_{k}\cdot {\bf r}_n + \theta_{k} + \alpha_{{\bf q}^{*}_{k}} \delta_{a{\rm II}},
\label{config_type1}
\end{eqnarray}
%q
where the $m_{xy}^2 \equiv \sum_{k=1}^{2} {{\Phi}_{\pm {\bf q}^{*}_k}^{+}}^2$ and $m_{z}^2 \equiv {{\Phi}_{\pm {\bf q}^{*}_3}^{+}}^2$, with the total amplitude $m$ given by $m^2=m_{xy}^2+m_{z}^2$, while $\theta_k$ represents a phase for each mode ${\bf \Phi}_{ {\bf q}^{*}_k}^{+}$. The $f_4$ of this double-$q$ (type 1) state is given by
\begin{eqnarray}
f_4^{\rm type1}  &=& \frac{9T}{20} \bigl[ \frac{3 m^4}{4} + m^2({{\Phi}_{\bf 0}^{-}}^2 - \frac{m_z^2}{2}) \nonumber \\
                 &+&  \frac{m_z^4}{2} + \frac{1}{2} {{\Phi}_{\bf 0}^{-}}^4 + 2{m_z}^2 {{\Phi}_{\bf 0}^{-}}^2 \nonumber \\
                 &+&  \frac{(m^2-m_{z}^2)}{\sqrt{2}}{\Phi}_{\bf 0}^{-} m_z 
                \{ {{\rm cos}\Theta+{\rm cos}(\Theta-\alpha^*)} \} \bigr] ,  \nonumber \\
               %  &&
\label{free_double_type1}
%m^2=8R^2+2R'^2,
\end{eqnarray}
where
\begin{eqnarray}
\Theta &\equiv& \theta_1+\theta_2+\theta_3,  \\
\alpha^* &\equiv& \alpha_{{\bf q}^{*}_{1}} +\alpha_{{\bf q}^{*}_{2}} + \alpha_{{\bf q}^{*}_{3}} .
\end{eqnarray}

After minimizing the obtained $f_4^{\rm type 1}$ with respect to the sum of the phase factor $\Theta$, we get
\begin{eqnarray}
f_4^{\rm type1} 
                 &=& \frac{9T}{20} \bigl[ \frac{3 m^4}{4} + m^2({{\Phi}_{\bf 0}^{-}}^2 - \frac{m_z^2}{2}) \nonumber \\
                 &+&  \frac{m_z^4}{2} + \frac{1}{2} {{\Phi}_{\bf 0}^{-}}^4 + 2{m_z}^2 {{\Phi}_{\bf 0}^{-}}^2 \nonumber \\
                 &-&  \frac{(m^2-m_{z}^2)}{\sqrt{2}}{\Phi}_{\bf 0}^{-} m_z 
                \sqrt{2+2 {\rm cos}\alpha^*}  \bigr],  \nonumber \\
\label{free_double_type1v2}
\end{eqnarray}
where $\Theta$ is determined as
\begin{eqnarray}
\Theta &=& \frac{\alpha^*}{2} + \pi.
\label{sum_phase}
\end{eqnarray}
In eq.~(\ref{free_double_type1v2}), $m$, $m_z$ and ${\Phi}_{\bf 0}^{-}$ are the variational parameters to be determined to minimize the total free energy for a given set of ($T$, $H$, $J_1/J_2$).

Note that this form of $f_4^{\rm type 1}$ contains the sublattice degree of freedom as the sum of sublattice phase differences $\alpha^*=\alpha_{{\bf q}^{*}_{1}} +\alpha_{{\bf q}^{*}_{2}} + \alpha_{{\bf q}^{*}_{3}}$, where each $\alpha_{{\bf q}^{*}_{k}}$ is determined by $J_1$ and $J_2$ via eq.~(\ref{eqalpha}) and by the direction of the critical wavevector (\textcolor{black}{see also Fig.~\ref{sum_of_alpha_NN2}(b)}). This should be contrasted with the triangular-lattice case where the lattice is Bravais without a sublattice structure.

We find that, if the direction of the critical wavevectors ${\bf q}^{*}_{1,2,3}$ is along the NN direction, the sum of the phase differences deviates from $\pi$ by a nonzero amount whose value is dependent on the $J_1$ and $J_2$ values, while, if the direction of the critical wavevectors is along the NNN direction, it is kept equal to $\pi$. In other words, the last term of eq.~(\ref{free_double_type1v2}) vanishes in the NNN case, while it yields a negative contribution in the NN case lowering the free energy. Therefore, even at the MF-level, the NN direction is preferred to the NNN direction for the critical wavevectors. This might explain the reason why the double-$q$ (type 1) state observed in MC simulation has wavevectors running along NN directions, as shown in \textcolor{black}{Figs.~11(a) and (b) in the main text.}

 We now move to the double-$q$ (type 2) state. The real-space sublattice spin configuration is given by
\begin{eqnarray}
\langle {\bf S}_{n}^{a} \rangle &\propto& 
   \left(
    \begin{array}{cc}
     \frac{m_{xy}}{2} \bigl[ {\rm cos} (\Delta_{n,1}) +    {\rm cos} (\Delta_{n,2})  \bigr] \\
     0 \\
     \frac{m_z}{\sqrt{2}} {\rm cos} (\Delta_{n,3}) +  \Phi_{\bf 0}^{-}
    \end{array}
  \right) , 
\label{config_type2}
\end{eqnarray}
and all configurations obtained by rotating it around the $z$-axis. The corresponding $f_4$ is given by
\begin{eqnarray}
f_4^{\rm type2}  %\nonumber \\
                 &=& \frac{9T}{20} \bigl[ \frac{9 m^4}{8} - \frac{5}{4} m^2 m_z^2 + \frac{7}{8} m_z^4 \nonumber \\
                 &+& \frac{1}{2} {{\Phi}_{\bf 0}^{-}}^4 + m^2 {{\Phi}_{\bf 0}^{-}}^2 +2m_z^2 {{\Phi}_{\bf 0}^{-}}^2 \nonumber \\
                 &+&  \frac{(m^2-m_{z}^2)}{\sqrt{2}}{\Phi}_{\bf 0}^{-} m_z \{{{\rm cos}\Theta+{\rm cos}(\Theta-\alpha^*)}\}   \bigr]. \nonumber \\ 
                 &&
\label{free_double_type2}
%m^2=4R^2+2R'^2,
\end{eqnarray}
After minimizing the obtained $f_4^{\rm type 2}$ with respect to $\Theta$, we get
\begin{eqnarray}
f_4^{\rm type2}  %\nonumber \\
                 &=& \frac{9T}{20} \bigl[ \frac{9 m^4}{8} - \frac{5}{4} m^2 m_z^2 + \frac{7}{8} m_z^4 \nonumber \\
                 &+& \frac{1}{2} {{\Phi}_{\bf 0}^{-}}^4 + m^2 {{\Phi}_{\bf 0}^{-}}^2 +2m_z^2 {{\Phi}_{\bf 0}^{-}}^2 \nonumber \\
                 &-& \frac{(m^2-m_{z}^2)}{\sqrt{2}}{\Phi}_{\bf 0}^{-} m_z  \sqrt{2+2 {\rm cos}\alpha^*} \bigr], \nonumber \\ 
                 &&
\label{free_double_type2v2}
\end{eqnarray}
where $\Theta$ is determined by eq.~(\ref{sum_phase}).

As in the case of the double-$q$ (type 1) state, the NN direction is preferred to the NNN direction for the critical wavevectors. Again, this is consistent with our MC observation.

\subsection{\label{sec:The double-$q$ (type 3) state}C. The double-$q$ (type 3) state}
In addition to the type 1 and 2 double-$q$ states, we find still another type of double-$q$ state, which we call ``type 3'' double-$q$ state. The state corresponds to the double-$q$ (type 3) state observed in our MC simulation. In the double-$q$ (type 3) state observed in MC simulations, the critical wavevectors turn out to differ between in the $xy$- and in the $z$-components. There exist four critical modes, {\it i.e.\/},  ${\bf \Phi}_{\pm{\bf q}^{*}_{1}}^{+}$ and ${\bf \Phi}_{\pm{\bf q}^{*}_{2}}^{+}$ in the $xy$ components, and ${\bf \Phi}_{\pm{\bf q}^{*}_{3}}^{+}$ and ${\bf \Phi}_{\pm{\bf q}^{*}_{4}}^{+}$ in the $z$-component. 

 With this MC observation in mind, we search for the MF solution with the four critical wavevectors, $q_1^*\sim q_4^*$. Although in the double-$q$ (type 3) state observed in MC simulations the critical wavevectors deviate even from the degenerate ring, we assume here that $q_1^*\sim q_4^*$ are wavevectors on the degenerate ring, which is required from the MF assumption. \textcolor{black}{With taking our MC results in Figs.~11(e)(f) into consideration, we also use assumption, ${\bf q}_{1}+{\bf q}_{2}+{\bf q}_{3}+{\bf q}_{4} \neq {\bf 0}$ in our MF analysis for the double-$q$ (type 3) state.}

The associated real-space sublattice spin configuration is given by
\begin{eqnarray}
\langle {\bf S}_{n}^{a} \rangle &\propto& 
   \left(
    \begin{array}{cc}
     \frac{m_{xy}}{2}    {\rm cos} (\Delta_{n,1}) \\
     \frac{m_{xy}}{2}    {\rm cos} (\Delta_{n,2}) \\
      \frac{m_z}{2} ({\rm cos} (\Delta_{n,3})+{\rm cos} (\Delta_{n,4})) + \Phi_{\bf 0}^{-} 
    \end{array}
  \right) , \nonumber \\
\label{config_type3}
\end{eqnarray}
and all configurations obtained by rotating it around the $z$-axis, where the amplitude $m_{xy}$ and $m_z$ are given by $m_{xy}^2 \equiv \sum_{k=1}^{2} {{\Phi}_{\pm {\bf q}^{*}_k}^{+}}^2$ and $m_{z}^2 \equiv \sum_{k=3}^{4} {{\Phi}_{\pm {\bf q}^{*}_k}^{+}}^2$. The total amplitude $m$ is given by $m^2=m_{xy}^2+m_z^2$.

 The corresponding $f_4$ is given by
\begin{eqnarray}
f_4^{\rm type3}
 &=& \frac{9T}{20} \bigl[ \frac{5 m^4}{8} +\frac{3 m_z^4}{4} -\frac{ m^2 m_z^2}{4} + \frac{1}{2} {\Phi_{\bf 0}^{-}}^4 \nonumber \\
 &+& (m^2+2m_z^2) {{\Phi}_{\bf 0}^{-}}^2 \bigr].
\label{free_double_type3}
\end{eqnarray}
The obtained $f_4$ does not depend on the sublattice degrees of freedom $\alpha_{{\bf q}^{*}_{i}}$, without any condition imposed between the $\theta$- and $\alpha$-variables unlike the double-$q$ (type 1) and (type 2) states. \textcolor{black}{Hence, in this double-$q$ (type 3) state, in sharp contrast to the double-$q$ (type 1 and type 2) states, there is no reason to select the higher symmetric NN or NNN directions for the critical wavevectors. Therefore, even possible selection of the intermediate direction may occur. Indeed, the ordering wavevectors observed in MC simulations slightly deviate from the NN direction in the double-$q$ (type3) state, which is consistent with the present MF result.}

\subsection{\label{sec:The collinear state}D. The triple-$q$ (collinear) state}
Our MC simulation has yielded the collinear triple-$q$ state. We find that such a state is also possible in the MF analysis. It consists of three individual critical modes ${\bf \Phi}_{\pm{\bf q}^{*}_{1}}^{+}$, ${\bf \Phi}_{\pm{\bf q}^{*}_{2}}^{+}$ and ${\bf \Phi}_{\pm{\bf q}^{*}_{3}}^{+}$ only in the $z$ component, whose amplitudes are equal to each other. The real-space sublattice spin configuration is given by
%
%\begin{eqnarray}
%\langle {\bf S}_{n}^{a} \rangle &\propto& 
%   \left(
%    \begin{array}{cc}
%      0 \\
%      0 \\
%      \frac{m}{2\sqrt{3}}    \sum_{k=1}^{3}\bigl[ {\rm cos} (\Delta_{n,k})-{\rm sin}(\Delta_{n,k}) \bigr]+\Phi_{\bf 0}^{-} 
%    \end{array}
%  \right) , \nonumber \\
%\label{config_collinear}
%\end{eqnarray}
%
\begin{eqnarray}
\langle {\bf S}_{n}^{a} \rangle &\propto& 
   \left(
    \begin{array}{cc}
      0 \\
      0 \\
      \frac{m}{\sqrt{6}}    \sum_{k=1}^{3} {\rm cos} (\Delta_{n,k} )+\Phi_{\bf 0}^{-} 
    \end{array}
  \right) , \nonumber \\
\label{config_collinear}
\end{eqnarray}
where the amplitude $m$ is given by $m^2 \equiv \sum_{k=1}^{3} {{\Phi}_{\pm {\bf q}^{*}_k}^{+}}^2$. We confirm that the spin texture described by eq.~(\ref{config_collinear}) actually represents the observed triple-$q$ collinear spin configuration shown in \textcolor{black}{Figs.~17 of the main text.} The corresponding $f_4$ is given by
\begin{eqnarray}
&f_4^{\rm collinear}&  \nonumber \\
    &=& \frac{9T}{20} \bigl[ \frac{5 m^4}{4} + \frac{1}{2} {\Phi_{\bf 0}^{-}}^4 + 3m^2 {{\Phi}_{\bf 0}^{-}}^2 \nonumber \\
    &+& \frac{\sqrt{6}}{3}m^3{\Phi_{\bf 0}^{-}}  
    \times \{ {{\rm cos}\Theta+{\rm cos}(\Theta -\alpha^*)} \}   \bigr]. \nonumber \\
\label{free_collinear}
\end{eqnarray}
Minimization of $f_4^{\rm collinear}$ with respect to  $\Theta$ yields,
\begin{eqnarray}
f_4^{\rm collinear}  %\nonumber \\
    &=& \frac{9T}{20} \bigl[ \frac{5 m^4}{4} + \frac{1}{2} {\Phi_{\bf 0}^{-}}^4 + 3m^2 {{\Phi}_{\bf 0}^{-}}^2  \nonumber \\
    &-& \frac{\sqrt{6}}{3}m^3{\Phi_{\bf 0}^{-}}   \sqrt{2+2 {\rm cos}\alpha^*} \bigr],\nonumber \\
\label{free_collinearv2}
\end{eqnarray}
where $\Theta$ is determined by the eq.~(\ref{sum_phase}).
In any case, for the collinear triple-$q$ state, the NN direction is preferred to the NNN one even at the MF-level as in the case of double-$q$ (type 1) and double-$q$ (type 2) states, which is consistent with our MC observation.
 \textcolor{black}{We can also understand the difference between the spin configurations on the two sublattices as observed in \textcolor{black}{Figs.~17(a) and (b) of the main text} at the MF level via the phase differences $\alpha_{{\bf q}^*_{k=1,2,3, {\rm NN}}}$.}

\subsection{\label{sec:The triple-$q$ state}E. The triple-$q$ (skyrmion-lattice) state}

We consider here the skyrmion-lattice triple-$q$ state which was observed to be stabilized in the triangular model\cite{Okubo2_s}. Although our MC simulation has indicated that the skyrmion-lattice triple-$q$ state \cite{Okubo2_s} is not stabilized in the present honeycomb model, we attempt a MF analysis in search for such a state. The state is defined as a superposition of the three individual critical modes ${\bf \Phi}_{\pm{\bf q}^{*}_{1}}^{+}$, ${\bf \Phi}_{\pm{\bf q}^{*}_{2}}^{+}$ and ${\bf \Phi}_{\pm{\bf q}^{*}_{3}}^{+}$ of equal weights, with an additional uniform component. We find the following real-space sublattice spin configuration, 
\begin{eqnarray}
\langle {\bf S}_{n}^{a} \rangle &\propto&  
       \frac{m_{xy}}{2\sqrt{3}} \sum_{k=1}^{3}{\rm sin} (\Delta_{n,k}) {\bf e}_{k}  \nonumber \\
    &+& \frac{m_z}{2\sqrt{3}} \sum_{k=1}^{3}{\rm cos} (\Delta_{n,k})  {\bf e}_z +\Phi_{\bf 0}^{-}  {\bf e}_z , \nonumber \\
\label{config_skyrmion}
\end{eqnarray}
where ${\bf e}_{k}$'s are unit vectors in the $xy$ plane which are summed up to zero as $\sum_{k=1}^{3} {\bf e}_{k}={\bf 0}$.
The amplitude $m_{xy}$ and $m_z$ are given by $m_{xy}^2 \equiv \sum_{k=1}^{3} {{\Phi}_{\pm k}^{xy}}^2$ and $m_{z}^2 \equiv \sum_{k=1}^{3} {{\Phi}_{\pm k}^{z}}^2$, where the ${\Phi}_{\pm k}^{xy(z)}$ is defined as the $xy$($z$) component of the critical mode ${\bf \Phi}_{\pm{\bf q}^{*}_{k}}^{+}$. Therefore, the triple-$q$ (collinear) state can be obtained as the case of the $m_{xy}=0$ limit. The corresponding $f_4$ is given by
\begin{eqnarray}
f_4^{\rm skyrmion} &=& \frac{9T}{20} \bigl[ \frac{3 m^4}{4} +\frac{2 m^2 m_z^2}{3} +\frac{7 m_z^4}{6}+(m^2+2 m_z^2){{\Phi}_{\bf 0}^{-}}^2 \nonumber \\
                   &+& \frac{1}{2} {{\Phi}_{\bf 0}^{-}}^4 +\frac{m_z (m^2+m_z^2) {\Phi}_{\bf 0}^{-}}{4\sqrt{6}} \{ {{\rm cos}\Theta+{\rm cos}(\Theta-\alpha^*)} \} \bigr], \nonumber \\
                 &&
\label{free_triple}
\end{eqnarray}
where $m^2$=$m_{xy}^2+m_z^2$. This equation can also be minimized with respect to $\Theta$, leading to 
\begin{eqnarray}
f_4^{\rm skyrmion} &=& {\frac{9T}{20} \bigl[ \frac{3 m^4}{4} +\frac{2 m^2 m_z^2}{3} +\frac{7 m_z^4}{6} +  (m^2+2 m_z^2){{\Phi}_{\bf 0}^{-}}^2 } \nonumber \\
                   &+& \frac{1}{2}{{\Phi}_{\bf 0}^{-}}^4 -\frac{m_z (m^2+m_z^2) {\Phi}_{\bf 0}^{-}}{4\sqrt{6}} \sqrt{2+2 {\rm cos}\alpha^*}  \bigr], \nonumber \\
                 &&
\label{free_triplev2}
\end{eqnarray}
where $\Theta$ is determined as in eq.~(\ref{sum_phase}) and is dependent of the sublattice structure. Hence, the free energy for the triple-$q$ state also depends on the direction of the critical wavevectors ${\bf q}^{*}_{1,2,3}$ as was discussed in~\textcolor{black}{subsection B} for the double-$q$ states. In fact, the free energy of the NN direction becomes lower than that of the NNN direction.

 The real-space sublattice spin configuration in the triple-$q$ (skyrmion-lattice) state of the present honeycomb-lattice model is slightly different from that in the triangular-lattice model\cite{Okubo2_s} in that the sum of the phase factors $\Theta$ is not strictly equal to $\pi$\cite{Okubo2_s}. However, we confirm that the total scalar chirality of the spin texture obtained from eq.~(\ref{config_skyrmion}) has a nonzero value even for $\Theta \ne \pi$~\cite{value_s}, and the resulting spin texture covers a whole sphere in the spin space as long as the relevant wavevectors are incommensurate with the lattice. Therefore, the real spin configuration obtained by eq.~(\ref{config_skyrmion}) may be called a skyrmion (or antiskyrmion) lattice. As mentioned, however, although this skyrmion-lattice state exists at the MF level \textcolor{black}{at least as a saddle-point solution}, it is not stabilized in the presence of fluctuations.

\subsection{\label{sec:The MF results}F. The relative stability of the MF phases}

\begin{figure}[h]
  \includegraphics[bb=200 50 612 590, width=5.0cm,angle=0]{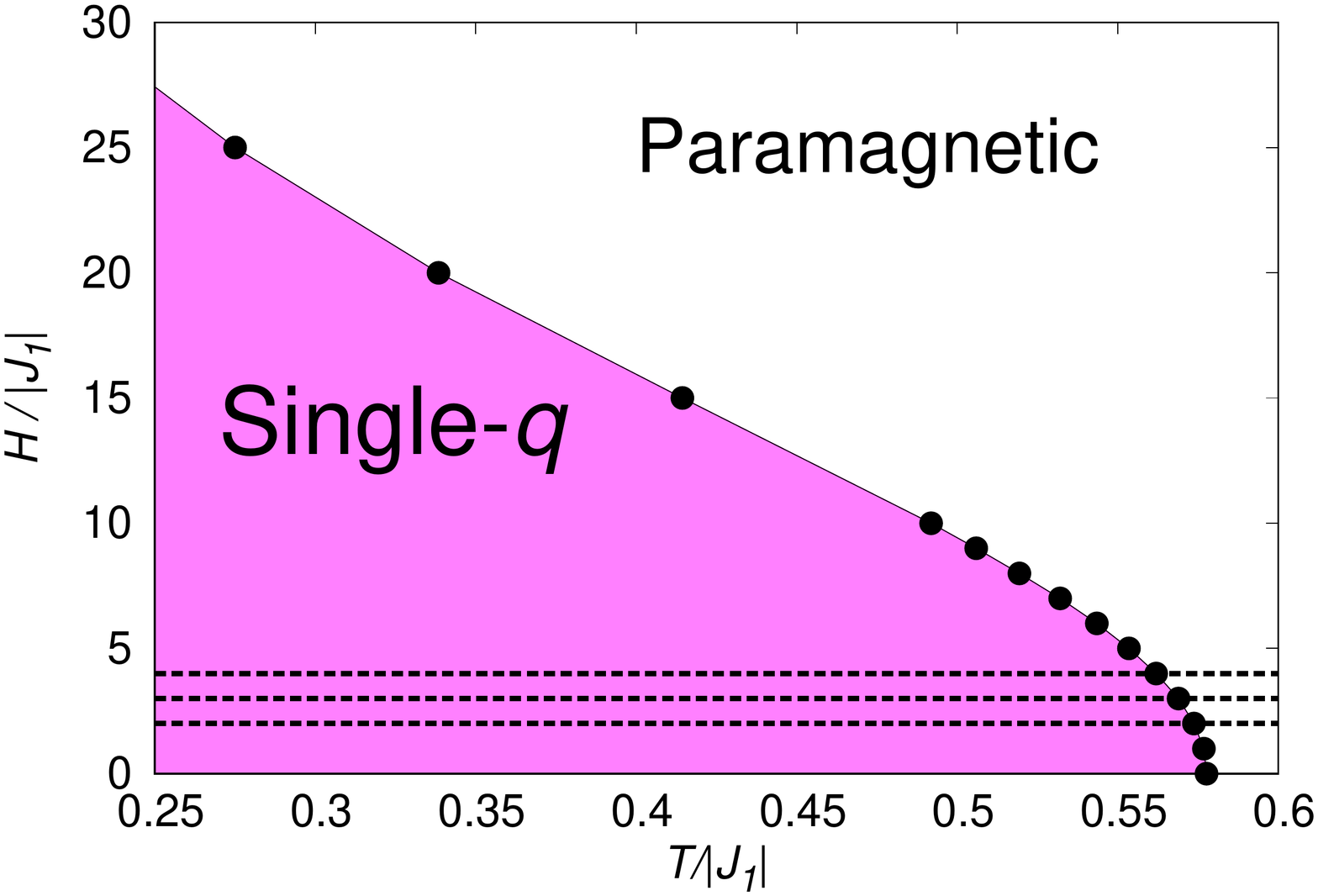}%-90->0
 \caption{(Color online) The MF phase diagram for $J_2/J_1=0.3$. The black dashed lines show the temperature regions at fixed fields for which the free energy differences are shown in Figs.~\ref{free_energy_J2_0.30}. 
 }
 \label{MF_phase_J2_0.30}
\end{figure}

 In this subsection, on the basis of the MF free energies of various multiple-$q$ states derived in the previous subsections, we examine the relative stability of various  multiple-$q$ phases at the MF level. In fact, for the case of $J_2/J_1=0.3$, the most stable state turns out to be the umbrella-type single-$q$ state as in the case of the triangular-lattice model~\cite{Okubo_s}.  \textcolor{black}{The resulting phase diagram in the temperature vs. magnetic field plane is given in  Fig.~\ref{MF_phase_J2_0.30}, where only the umbrella-type single-$q$ state appears. This result is in sharp contrast to the rich phase diagram determined by MC simulations shown in Fig.~3 of the main text.} The other point to be noticed might be that the energy scale associated with the ordering, which manifests itself as the transition temperature, is an order of magnitude greater than the actual transition temperature determined by MC simulations. 

\begin{figure}[t]
  \includegraphics[bb=100 50 612 792,width=11.5cm,angle=0]{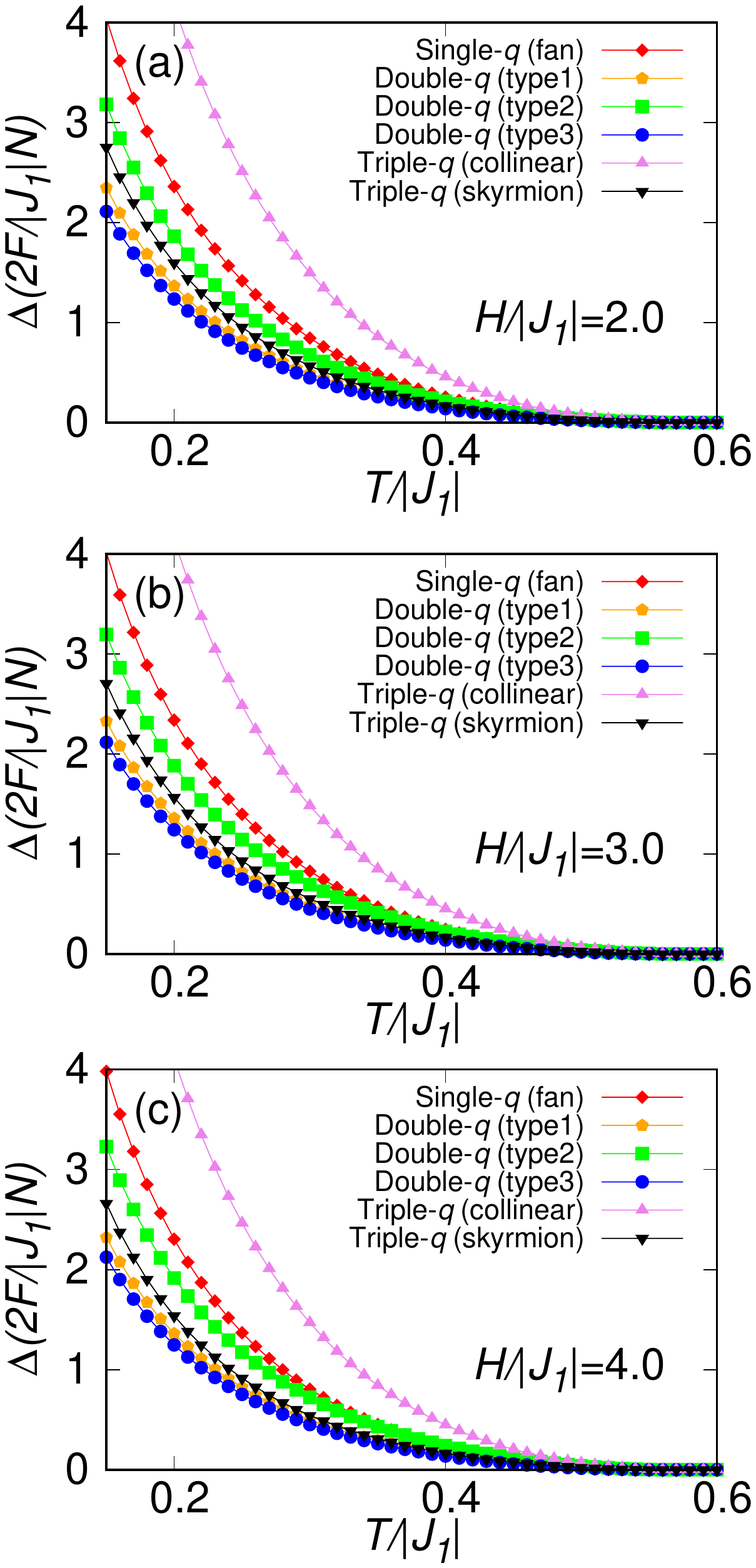}
 \caption{(Color online) The temperature dependence of the MF free-energy difference for $J_2/J_1=0.3$, where  \textcolor{black}{the free energy ($2F/|J_1|N$) of eq.~(\ref{free3}) with $f_4^{\rm umbrella}$ of eq.~(\ref{free_single})} is taken as the energy origin. The field intensities are \textcolor{black}{(a) $H/|J_1|=2.0$, (b) $H/|J_1|=3.0$, and (c) $H/|J_1|=4.0$}.
}
 \label{free_energy_J2_0.30}
\end{figure}

 In Figs.~\ref{free_energy_J2_0.30}, we show for several representative field intensities the temperature dependence of the computed  MF free energy of various multiple-$q$ states, \textcolor{black}{$\Delta$($2F/|J_1|N$)}, in which the free energy of the umbrella-type single-$q$ state \textcolor{black}{($2F/|J_1|N$) with $f_4^{\rm umbrella}$} is taken as the energy origin. As the umbrella-type single-$q$ state always has the lowest free energy,  \textcolor{black}{$\Delta$($2F/|J_1|N$)} is always non-negative. The associated energy scale of the ordering turns out to be of order unity, which is the cause of the transition temperature being significantly higher than that of MC simulation. Of course, we should be careful about the fact that the ordered states identified by finite-temperature MC simulations are not true long-range-ordered states, only quasi-long-range-ordered states, while the MF analysis treat these states as long-ranged-ordered states. This observation in turn highlights the importance of fluctuations in the ordering process of the model. 

 Yet, the MF calculation provides us some suggestive information; (i) At lower temperatures, the double-$q$ (type 3) state has the second lowest free energy for any fields. (ii) The double-$q$ (type 1) state is always competing with the double-$q$ (type 3) state, especially for intermediate magnetic fields. The properties (i) and (ii) might help us to understand why the double-$q$ (type 3) state occupies a relative wide region of the MC phase diagram of \textcolor{black}{Fig.~3 in the main text}, and the double-$q$ (type 1) state appears at intermediate magnetic fields. (iii) Strong magnetic fields are required to lower the free energy of the single-$q$ (fan) state. This property (iii) might explain the reason why the fan-type single-$q$ state is stabilized only at higher magnetic fields in the MC phase diagram. (iv) The double-$q$ (type 2) and the triple-$q$ (collinear) states are hard to be stabilized at lower temperatures since they always have free energies significantly higher than those of other states at lower temperatures. This property (iv) also seems to be consistent with our present MC observation.

\textcolor{black}{Although our MF analysis suggests that the skyrmion-lattice state is more stable than the states like the single-$q$ (fan), the double-$q$ (type2) and the triple-$q$ (collinear) states which turn out to be stabilized in a certain $T$ and $H$ region in the MC phase diagram of Fig.~3 in the main text, the triple-$q$ (skyrmion) state itself is never stabilized at any $T$ and $H$ in our MC phase diagram. We don't know the clear reason why the triple-$q$ (skyrmion) state is not stabilized in the present $J_1$-$J_2$ honeycomb-lattice model, in sharp contrast to the case of the $J_1$-$J_2$ triangular-lattice model where the triple-$q$ (skyrmion) state is stabilized over a certain range of $T$ and $H$~\cite{Okubo2_s}.
}

\end{document}